\documentclass[twocolumn]{aastex62}

\newcommand{\vperp}{v_{\rm loc}}
\newcommand{\kms}{\hbox{km\,s$^{-1}$}}
\newcommand{\teff}{$T_{\rm eff}\,$}
\newcommand{\msun}{M_\odot}
\newcommand\vsini{$v_r\sin i $}

\graphicspath{{./}}

\received{} 
\revised{} 
\accepted{} 
\submitjournal{ApJ}

\shorttitle{Dynamical vs Supernova Ejections}
\shortauthors{Dorigo Jones et al.}

\begin{document}
\title{Runaway OB Stars in the Small Magellanic Cloud: Dynamical Versus Supernova Ejections}

\correspondingauthor{M. S. Oey}
\email{msoey@umich.edu}

\author{J. Dorigo Jones}
\affiliation{Department of Astronomy, University of Michigan, 1085 South University Ave., Ann Arbor, MI 48109-1107, USA}

\author{M. S. Oey}
\affiliation{Department of Astronomy, University of Michigan, 1085 South University Ave., Ann Arbor, MI 48109-1107, USA}

\author{K. Paggeot}
\affiliation{Department of Astronomy, University of Michigan, 1085 South University Ave., Ann Arbor, MI 48109-1107, USA}

\author{N. Castro}
\affiliation{Leibniz-Institut f\"ur Astrophysik, An der Sternwarte 16, 14482 Potsdam, Germany}

\author{M. Moe}
\affiliation{Steward Observatory, University of Arizona, 933 N. Cherry Ave., Tucson, AZ 85721, USA}

\begin{abstract}
Runaway OB stars are ejected from their parent clusters via two mechanisms, both involving multiple stars: the dynamical ejection scenario (DES) and the binary supernova scenario (BSS). We constrain the relative contributions from these two ejection mechanisms in the Small Magellanic Cloud (SMC) using data for 304 field OB stars from the spatially complete, Runaways and Isolated O-Type Star Spectroscopic Survey of the SMC (RIOTS4). We obtain stellar masses and projected rotational velocities \vsini\ for the sample using RIOTS4 spectra, and use transverse velocities $\vperp$ from {\it Gaia} DR2 proper motions. Kinematic analyses of the masses, \vsini, non-compact binaries, high-mass X-ray binaries, and Oe/Be stars largely support predictions for the statistical properties of the DES and BSS populations. We find that dynamical ejections dominate over supernova ejections by a factor of $\sim 2-3$ in the SMC, and our results suggest a high frequency of DES runaways and binary ejections. Objects seen as BSS runaways also include two-step ejections of binaries that are reaccelerated by SN kicks. We find that two-step runaways likely dominate the BSS runaway population. Our results further imply that any contribution from \textit{in-situ} field OB star formation is small. Finally, our data strongly support the post-mass-transfer model for the origin of classical Oe/Be stars, providing a simple explanation for the bimodality in the \vsini\ distribution and high, near-critical, Oe/Be rotation velocities. The close correspondence of Oe/Be stars with BSS predictions implies that the emission-line disks are long-lived.
\end{abstract}


\section{Introduction} \label{sec:intro}

Given their relatively short lifetimes, most massive stars are located in the clusters or OB associations in which they formed, with the stars having all condensed from the same giant molecular cloud (GMC) \citep[e.g.,][]{LadaLada03}. However, studies show that there are many OB stars that do not belong to any cluster or OB association. Indeed, about 20 -- 30\% of all massive stars are isolated from other massive companions and are thus said to exist in the ``field,'' the sparsely populated region outside of star clusters \citep[e.g.,][]{Gies87, Oey04, deWit05, Lamb16}. For decades, the existence of massive field stars has both challenged and advanced our understanding of stellar kinematics. Now, following the release of {\it Gaia} DR2 proper motions, we are able to shed light on this topic and to ultimately probe the kinematic evolution of massive stars.

One possibility for the origin of field OB stars is that some of them could actually form outside of OB associations, and could therefore form \textit{in situ} in the field \citep{deWit05, Parker07, Oey2013}. This challenges theories on massive star formation that state the necessary gas conditions occur only in dense regions of GMCs \citep{McKee07}. The other possibility, and the one that we examine in this work, is that OB stars are ejected from their birth clusters into the field as runaway stars, which are traditionally defined to have peculiar motion of at least 30 $\kms$ \citep[e.g.,][]{Gies87, Leonard88, Clarke92, Hoogerwerf00}. 

Runaway OB stars can be produced via two possible ejection mechanisms, both involving multiple stars: (1) the dynamical ejection scenario (DES), in which a close 3- or 4-body interaction in a dense cluster core ejects a massive star \citep{Blaauw61, Poveda67, Leonard88}; and (2) the binary supernova scenario (BSS), in which the core-collapse supernova (SN) of the more evolved star in a massive binary system catapults the OB companion into the field \citep{Zwicky, Blaauw61, Heuvel81}. In the BSS, the SN explosion ejects mass from the system and induces a natal kick on the newly-formed compact object that can disrupt the system, ejecting the OB companion at a space velocity comparable to its final pre-SN orbital velocity \citep{Blaauw61, Renzo19}. If the system remains bound post-SN, the system may be observable as a high-mass X-ray binary (HMXB) \citep{Gott71}. The relative contributions from the two ejection mechanisms are still poorly understood. This is partly due to the lack of statistically complete data on their kinematics, a situation that is being remedied by {\it Gaia}.

It is well-established that both ejection mechanisms require massive binaries in order to produce runaway OB stars. In the BSS, the SN explosion of the more evolved star in a massive binary results in the ejection of either the single OB companion or the entire system \citep{Renzo19}. In the DES, due to gravitational focusing, the majority of runaway stars are produced via interactions with massive binaries \citep{PeretsSubr}. In addition, it is predicted that a single very massive ``bully'' binary at the center of the cluster dominates the cross section for interaction and ejects stars via close gravitational encounters until it is ejected itself \citep{FujiiPZ}.  Furthermore, the properties of massive binary stars in clusters, such as their periods and separations, initial mass ratios, and eccentricities, can affect the velocities and multiplicities of runaways produced by both ejection mechanisms \citep[e.g.,][]{OhKroupa}.

In \citet[][hereafter Paper~I]{PaperI}, we argued that dynamical ejections dominate over SN ejections for the field OB runaways in the Small Magellanic Cloud (SMC).  This was based on our observations that eclipsing binaries (EBs) and double-lined spectroscopic binaries (SB2s), which are both tracers of the DES, reach much higher velocities and more closely match the total distribution than our high-mass X-ray binaries do. We also note that half of our field OB stars have a transverse proper motion greater than 39 $\kms$, a speed that unbound companions from SN ejections are rarely predicted to reach \citep[e.g.,][]{Renzo19}. This therefore suggests that the majority of our stars at runaway velocities are produced via the DES.

In this paper, we perform a comprehensive follow-up analysis of the preliminary work presented in Paper~I, which exploits {\it Gaia} data for field OB stars in the SMC. Our sample consists of 304 SMC field OB stars observed in the spatially complete RIOTS4 survey \citep{Lamb16}. We use (1) local residual transverse velocities ($\vperp$) calculated from the {\it Gaia} DR2 proper motions, (2) the stellar masses, and (3) measured projected rotational velocities (\vsini) of our stars to more quantitatively constrain the relative contributions from the two ejection mechanisms and to ultimately learn more about the kinematic evolution of massive stars.  Together with Paper~I, we provide the first kinematic analysis of a statistically complete sample of field OB stars in an external galaxy, which can be used to test predictions for the DES and BSS.  

In Section~\ref{sec:intro}, we outline the background and theoretical expectations for the DES and BSS. In Section~\ref{sec:distribution}, we present our sample of SMC field OB stars, and discuss their kinematics and binary statistics.  We estimate the runwaway frequencies produced via the DES, and independently, the BSS, based on theoretical predictions applied to our sample. In Section~\ref{sec:mass}, we present the stellar masses and discuss the kinematics in terms of both mass and velocity. In Section~\ref{sec:rotation}, we present our projected rotational velocities \vsini, and discuss the kinematics in terms of \vsini\ and transverse velocity.  We argue that the number of classical Oe/Be stars may be a useful surrogate for BSS ejections, which offers further constraints on the fraction of BSS runaways. In Section~\ref{sec:discussion}, we explore the consequences of our results, in particular, the significance of two-step ejections in our sample, the ratio of DES to BSS runaways, and the origin of Oe/Be stars.

\subsection{Dynamical vs SN Ejection} \label{subsec:Predictions}

Regarding the dynamical ejection scenario, there are many numerical studies in the literature. The main driver for generating massive runaways is the formation of a very dense central core of stars due to gravothermal collapse, which drives more massive stars toward the center \citep{FujiiPZ, OhKroupa}. This greatly increases the probability of close encounters that can slingshot stars into the field. The cluster core collapse is halted only by the energy from hard binaries, either newly-formed or primordial, which act as a kinetic energy source for the core \citep{Fujii14}.

\citet{FujiiPZ} conducted N-body simulations with a range of cluster masses from 6,300 $M_{\odot}$ to 200,000 $M_{\odot}$, for a Salpeter initial mass function (IMF) and fixed central density. Their simulations do not include primordial binaries, although they do consider binaries that form dynamically. They find that stellar ejections are strongly dominated by a ``bully'' binary located in the core, that is relatively wide and composed of the most massive stars in the cluster, yielding the greatest interaction cross section. The bully binary is naturally produced via the gravothermal collapse of the cluster core, which occurs within about 1 Myr for the clusters simulated. Since the bully binary dominates the cross section for interactions, the encounter probability depends only weakly on the masses of the other, single stars. Therefore, the mass distribution of the runaway OB stars is not strongly modified; they find that the mass distribution for low-mass runaway stars $< 8 M_{\odot}$ is consistent with the Salpeter slope.  However, massive runaway stars $> 8 M_{\odot}$ are significantly over represented, in particular for runaways $> 30 M_{\odot}$. In addition, \citet{FujiiPZ} examine the mean escape velocity as a function of mass for the runaways produced after 3 Myr, which gives more than enough time for the bully binary to form. They find a relatively constant median velocity near $42 \ \kms$ for ejected stars $> 20 M_{\odot}$.

\citet{PeretsSubr} carry out N-body simulations in which all massive stars reside in primordial binaries.  They model a 5,000 $M_{\odot}$ cluster with stellar masses ranging from 0.2 $M_{\odot}$ to 80 $M_{\odot}$ following a Salpeter IMF, and the binaries have initially zero eccentricity and semi-major axes set between 0.05 AU and 50 AU. The cluster is evolved to an age of 2.7 Myr. They find that the velocity distribution of escaping stars having velocities in the range $20 \ \kms$ to $150 \ \kms$ is independent of binary separations and cannot be produced via single-single encounters alone, further suggesting the presence of the bully binary that dominates the interaction cross section, in agreement with \citet{FujiiPZ}. \citet{PeretsSubr} find that the runaway fraction increases with mass, with the O star runaway fraction being two to three times greater than that of B stars, and the more massive stars also have higher ejection velocities. Furthermore, the DES can also generate runaway, non-compact OB binaries. The binary fraction of DES runaways decreases with ejection velocity, falling from $\sim$40\% at velocities of 30 $\kms$ to $\sim$10\% at 150 $\kms$.

\citet{OhKroupa} carry out comprehensive simulations of $10^{3.5}\ \msun$ clusters evolved to 3.5 Myr age, studying the effects of initial conditions including mass segregation, binary fraction, period distribution, binary mass ratios, and eccentricities.  They find that high stellar density is the dominant parameter driving high runaway frequencies, which is also aided by high primordial binary fractions, since ejections happen early, peaking around ages of $\sim$1 Myr. As also found by others \citep[e.g.,][]{FujiiPZ, PeretsSubr}, O stars typically are ejected more frequently than B stars, therefore causing the mass function of all ejected OB stars, including runaways and walkaways, to be flatter than the parent IMF.  However, the frequencies of only runaway O- vs B-stars may be more similar, depending on the mass ratios and period distribution; in particular, tight pairs of O-star ``twins'' are the most dynamically active and produce the fastest runaways.  In general, O-stars tend to have faster runaway velocities, in some cases exceeding 200 $\kms$ \citep[e.g.,][]{PeretsSubr}. \citet{OhKroupa} note that the peak of the velocity distribution is closely related to the cluster mass and density, since the cluster potential determines the escape velocities. Thus, accounting for the cluster mass distribution would act to weight the velocity distribution toward lower values \citep{OhKroupa15}. The ejected binary frequencies are typically around $\sim$30\%, and are biased toward short periods.

The binary supernova scenario is also well studied. For massive binaries, the core-collapse SN of the more evolved star results in one of two outcomes: (1) an unbound OB companion with an ejection velocity similar to its final pre-SN orbital velocity plus a poorly constrained SN kick; or (2) a bound binary consisting of the newly-formed compact object and an OB companion, possibly observable as an HMXB.  The reversal of the mass ratio prior to the explosion and widening of the orbit tend to inhibit acceleration to runaway velocities ($v > 30\ \kms$). \citet{Brandt95} carried out simulations of massive binary systems to evaluate the post-SN binary and HMXB properties. They modeled the outcomes for a binary system of fixed initial stellar masses and a constant logarithmic distribution of initial orbital periods, with a SN kick velocity distribution based on the observed pulsar birth velocities. They demonstrate that the period anti-correlates with both kick velocity and kick directions opposing the orbital motion, resulting in a strong anti-correlation between the final orbital separation and post-SN systemic velocity. The bulk of their massive binaries that remain bound have periods of $< 100$ days, which also is roughly the threshold for 30 $\kms$ runaway velocities. About 1/4 of these systems obtain eccentricities leading to strong dynamical mass transfer, inducing merging or disruption; thus ultimately about 20\% of their massive binaries remain stably bound. They also note that the remaining star's spin constitutes a substantial fraction of the total orbital angular momentum, implying high rotation velocities.

More recently, \citet{Renzo19} perform comprehensive binary population synthesis modeling to study populations of BSS products, including unbound single-stars. The binaries are mainly characterized by three parameters: (1) the primary-star masses, ranging from 7.5 $M_{\odot}$ to 100 $M_{\odot}$ and weighted by a Kroupa IMF; (2) a flat distribution in initial mass ratio $q$, ranging from 0.1 to 1; and (3) an empirically motivated, --0.55 power-law distribution for the initial orbital period over a range of log(0.15/days) to log(5.5/days). In the fiducial model they assume a metallicity of $Z = 0.02$; however, they also consider lower metallicities. They also treat a variety of evolutionary processes including wind mass loss, mass-transfer efficiency, common envelope evolution, and SN kick parameters. 

The results of \citet{Renzo19} are generally consistent with those of \citet{Brandt95} for bound runaways. For the unbound OB stars, they find that in general the ejection velocities are too slow to produce large numbers of runaways with velocities $> 30\ \kms$. The mean ejection velocity for their fiducial population of OB companions is $\sim$12 $\kms$, and toward higher companion masses, the mean ejection velocities are even lower. This is indeed the opposite trend that we expect from the DES as discussed above. This robust prediction for low runaway frequencies is partly due to the fact that essentially all massive runaways have experienced mass transfer such that the secondary becomes more massive than the primary, widening the orbit before the primary explodes. Thus, both the increased mass of the ejected star and the lower orbital velocity contribute to the slower ejection speeds. In their fiducial model at solar metallicity, 67\% of the OB companions are disrupted, such that only 2.5\% of the parent binaries are ultimately ejected as single runaways with $v > 30\ \kms$ and 48\% have $v < 30\ \kms$. These simulations give an 11\% frequency for post-SN bound systems, out of all OB binaries. The remaining 39\% of the parent binaries result in stellar mergers pre-SN (22\%) and evolved single runaways and walkaways (17\%). \citet{Kochanek19} also carried out binary population synthesis models that are largely consistent with these branching ratios.

We will use these predicted kinematic trends for the DES and BSS to discriminate between these two ejection mechanisms. A third, hybrid ``two-step'' mechanism also generates runaways that correspond to a small subset of both DES and BSS populations \citep{Pflamm-Altenburg10}. We will also evaluate the effect of these objects on the relative total contributions of each mechanism.

\section{OB Runaway Kinematics} \label{sec:distribution}

To evaluate the two ejection mechanisms, we consider the sample of SMC field OB stars from the Runaways and Isolated O-Type Star Spectroscopic Survey of the SMC \citep[RIOTS4;][]{Lamb16} that have reliable {\it Gaia} proper motions (PMs) from Paper~I.  There are slight differences in the default sample we use for analysis in the present work:  The sample here includes the 11 stars that we defined to exist in a ``Boundary'' group between the SMC Wing and Bar; and we drop from the sample the four objects in Paper~I with the highest PM, since we now understand these to be spurious measurements. The RIOTS4 field OB stars were selected to be those maximally isolated, at least 28 pc in projection away from any other OB candidate from the sample of \citet{Oey04}; however, seven stars that did not meet this criterion were inadvertently included in RIOTS4 and are now also deleted. This leaves a total of 304 stars in the present sample. Of these, 15 are eclipsing binaries (EBs; \citealt{Pawlak13}), 11 are double-lined spectroscopic binaries (SB2s; \citealt{Lamb16}), and 14 are high-mass X-ray binaries (HMXBs; \citealt{HaberlSturm}) (see Paper~I). Runaway EBs and SB2s trace dynamical ejections, while runaway HMXBs trace bound SN ejections, as mentioned in Section~\ref{sec:intro}.

We use the local residual transverse velocity, $\vperp$, from Paper~I\footnote{We clarify that in Table~1 of Paper~I,  columns 11 and 13 correspond to systemic RA and Dec velocities, respectively, of the local fields for each target star.}.  The values were obtained from {\it Gaia} DR2 PMs and corrected for each star's local velocity field within a 90-pc radius to obtain its residual transverse velocity. We note that the conventional definition of runaway stars specifies a 30 $\kms$ threshold for the 3-D space velocity, which corresponds to a transverse $\vperp \geq 24\ \kms$. Our median {\it Gaia} measurement error on $\vperp$ for this sample is 27 $\kms$, and so in this work, we consider runaways to have $\vperp \geq 30\ \kms$, which is equivalent to sampling stars with 3-D space velocities $\geq 37\ \kms$.

Of our 304 field OB stars, 197 stars, or $65\% \pm 6\%$, have $\vperp > 30\ \kms$. For comparison, we also find that 220 stars, or $72\% \pm 6\%$, have $\vperp \geq 27\ \kms$, our nominal detection limit. It is therefore apparent that less than half of our stars have $\vperp$ below the {\it Gaia} detection limit, implying a large runaway population (Paper~I), especially since $\vperp$ is a lower limit on the star's 3-D space velocity.

The remaining stars correspond to a slower population, which comprises 107 stars with $\vperp < 30\ \kms$, or $35\% \pm 4\%$ of the sample. These stars consist of ``walkaways,'' which we define as stars unbound from clusters at velocities below the runaway threshold, any field stars that formed \textit{in situ} \citep[e.g.,][]{Lamb10, Oey2013}, and runaways with trajectories oriented largely in the line of sight. Since our objects are selected to be at least 28 pc in projection from other OB stars, the sample is biased against walkaways. For illustration, a 20 $\kms$ walkaway star will travel 20 pc in 1 Myr. \citet{Renzo19} show that BSS walkaways travel average distances on the order of $\sim$60 pc for O stars at SMC metallicity.  Thus, allowing for walkaways to travel 28 pc to be included in our sample, our incompleteness for walkaways due to selection bias is on the order of 2.4, including projection effects (see Section~\ref{subsec:incompleteness}). DES walkaways suffer somewhat lower incompleteness, since DES ejections are on average faster and occur earlier \citep[e.g.,][]{OhKroupa}.  

Given that the velocity distribution peaks near the observational detection limit, it is therefore likely that our non-runaway population remains strongly dominated by walkaways, rather than objects that formed {\it in situ}. This is consistent with the fact that our search for small clusters surrounding our target stars yields very few candidates \citep{VargasSalazar20}. We shall show in this work that the frequencies of DES and BSS runaways are fully consistent with the contribution of {\it in-situ} stars being negligible (Section~\ref{subsec:insitu}). Since our sample of field stars represents $\sim$25\% of all SMC OB stars \citep{Oey04}, the data above imply that $>16\% \pm 4\%$ of OB stars are runaways ($\vperp > 30\ \kms$), which is consistent with the frequency of $\sim$20\% reported for the Milky Way \citep[e.g.,][]{Gies86, Stone91}.

\subsection{The Frequency of Dynamical Runaways} \label{subsec:DESpred}

\begin{figure}
\plotone{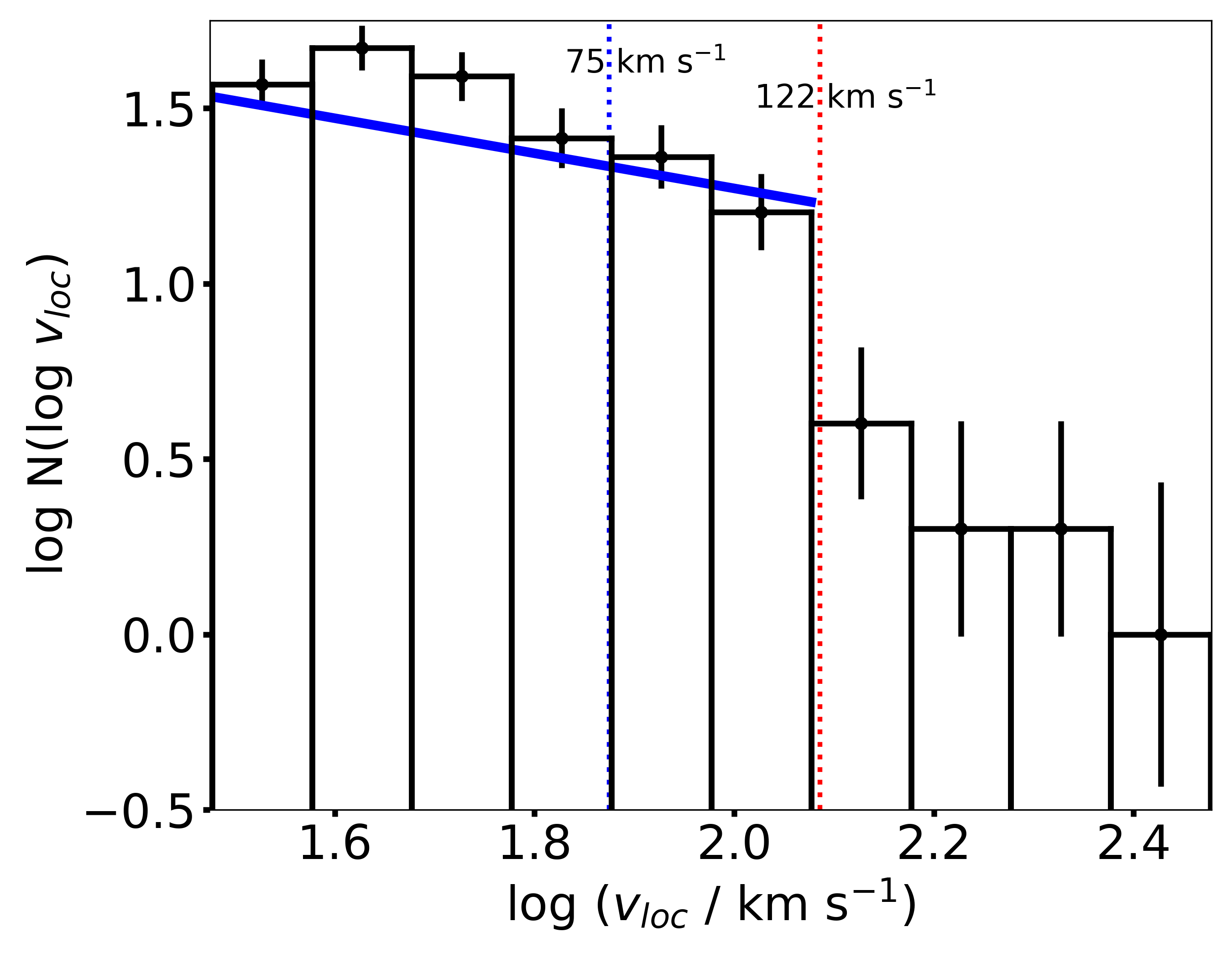}
\caption{Velocity distribution for the 197 RIOTS4 stars with runaway velocities ($\vperp > 30 \ \kms$), binned by 0.1 dex. Shown is the predicted $\vperp^{-1.5}$ relation from \citet{PeretsSubr}, which in $\log N(\log v)$ space has slope --0.5. The relation is normalized to the observed number of stars in the two bins between the indicated velocity range used for extrapolating the number of DES runaways. \label{f_PM_distrib}}
\end{figure}

We expect the highest-velocity stars to be due to dynamical ejections (Section~\ref{subsec:Predictions}). Based on their models described above, \citet{PeretsSubr} predict a runaway velocity distribution of $v^{-3/2}$ below $\sim$150 $\kms$, which projected to 2-D corresponds to $\vperp$ of $\sim$122 $\kms$, steepening to $v^{-8/3}$ at higher velocities. \citet{OhKroupa} find similar slopes of --1.4 to --2.1, fitted over the entire runaway range. These works do not consider the effects of steady-state star formation with a cluster mass function, but \citet{OhKroupa} find that O- and B-star velocity distributions are similar to each other at the highest velocities, and so the composite velocity distribution for runaways should be quite robust. In Figure~\ref{f_PM_distrib}, we see that the predicted velocity relation for $\vperp < 122 \ \kms$ from \citet{PeretsSubr} agrees well with our runaway velocity distribution, although we note that the observed distribution is somewhat steeper, as would be expected by significant contamination from BSS runaways at lower velocities (see Sections~\ref{subsec:PredBSS},~\ref{subsec:vsini_runaway}). Thus, the overall distribution is consistent with our finding in Paper~I that the DES mechanism dominates the runaway population in our sample.

Adopting the predicted $\vperp^{-3/2}$ distribution below 122 $\kms$, we can therefore roughly estimate the total number of dynamically ejected runaways out of our sample of 304 field OB stars by extrapolating from the high-velocity tail. Assuming that the 40 stars with $75\ \kms \leq \vperp < 122\ \kms$ are all dynamically ejected runaways, we obtain 108 DES runaways in the range $30 \ \kms \leq \vperp < 75\ \kms$. There are a total of 149 stars observed in this velocity range, therefore implying that 41 stars, or $21\% \pm 7\%$ are due to the BSS. Adding the 40 stars with $75\ \kms \leq \vperp < 122\ \kms$ and also the 8 very highest-velocity stars ($\vperp > 122\ \kms$), which are all assumed to be dynamically ejected, to the 108 predicted DES runaways in the range $30 \ \kms \leq \vperp < 75\ \kms$, we obtain a total DES contribution of $79\% \pm 8\%$ (= 156/197) to the runaway population.  Runaways comprise $65\% \pm 6\%$ (= 197/304) of our sample, and so DES runaways alone, without walkaways, are about half ($51\% \pm 7\%$) of all our SMC field OB stars. The errors quoted are for Poisson statistics only, and do not account for systematic uncertainties; for example, it remains possible that a few objects in the $75-122\ \kms$ velocity range are BSS runaways, implying a possible slight overestimate in our total DES frequency. One HMXB has $\vperp$ in this range, supporting this possibility (Section~\ref{subsec:twostep}).

\subsection{The Frequency of SN Runaways} \label{subsec:PredBSS}

We can also make an independent estimate of the contribution of BSS runaways to our sample. In a steady-state population with a constant star formation rate, the OB runaway fraction from BSS is smaller than from DES. \citet{Renzo19} find that 1.2\% and 10\% of OB stars with $m > 15\ M_{\odot}$ in a steady-state population at SMC metallicity are ejected single-star runaways and walkways, respectively. However, the post-SN bound frequency is 13\% for companions to black holes (BH) or neutron stars (NS), a large fraction of which achieve $v > 30\ \kms$, depending on the assumed kick velocities and BH fallback prescriptions. Thus, a total of roughly 24\% of OB stars in a steady-state population are the survivors of post-SN binary systems and the remaining $\sim$76\% are primaries that have not yet exploded. This is consistent with the findings of \citet{Moe17}, who estimated that $\sim$20\% of OB stars in a steady-state population are the secondaries in which the true primaries have exploded as SNe.

About half of the post-SN systems (13\%) remain bound. The models at SMC metallicity predict a total walkaway-to-runaway ratio of 11.1, so we therefore expect that $\sim$2\% of all OB stars are BSS runaways. The RIOTS4 field stars correspond to about 25\% of the SMC population \citep{Oey04}, and therefore we expect that about 8\% of our field stars are runaways due to the BSS. Similarly, since $65\% \pm 6\%$ of our field stars are runaways, then BSS runaways are about 12\% $\pm 2\%$ of runaways. This is slightly lower than our estimate of $21\% \pm 7\%$ based on the prediction for DES runaways (Section~\ref{subsec:DESpred}), and moreover, this breakdown does not account for objects that experience both DES and BSS runaway acceleration, including two-step runaways. We carry out a full accounting of the DES and BSS runaways below in Section~\ref{sec:discussion}.

Altogether, these first-order estimates suggest that the DES:BSS allocations are roughly in the range 80:20, respectively, among the SMC runaway OB stars. These values will be adjusted when we fully evaluate the ratio of DES to BSS runaways in Section~\ref{subsec:DESvsBSS}.

\subsection{Binary Runaways} \label{subsec:binary}

It is possible for the DES to eject massive binaries, in addition to single stars.  Therefore, non-compact binary runaways, such as EBs and SB2s, are a direct probe of the DES mechanism. Faster runaway binaries can be produced via binary-binary interactions in a dense cluster core; whereas binary interaction with a single star causes the binary typically to be ejected at much lower velocities \citep{FujiiPZ, PeretsSubr}. We are thus in a position to compare the characteristics of our non-compact, massive binary runaways to predictions, to better understand their production and interaction histories.  

\citet{PeretsSubr} find a maximum velocity of $\lesssim 200\ \kms$ for dynamically-ejected binary runaways in their simulations, while some models of \citet{OhKroupa} have maximum speeds that are double this value. Taken at face value, our highest-velocity binary is an EB (M2002-81258) with $\vperp = 201 \pm 36\ \kms$ and period of 2.7 days (\citealt{Pawlak16}). Two more high-velocity binaries are an SB2 (M2002-36213) with $\vperp = 121 \pm 32\ \kms$, and an EB \& SB2 star (M2002-65355) with $\vperp = 109 \pm 34\ \kms$ and period of 1.2 days (\citealt{Pawlak16}). These high-velocity non-compact binaries are likely the result of a close interaction with a hard and/or massive bully binary in the cluster core. Although we caution that individual {\it Gaia} proper motions may have unknown errors, the short periods are consistent with expectation for high-speed binaries. We find that in general, the binary frequency of our EBs and SB2s decreases with increasing velocity, a result consistent with the simulations of \citet{PeretsSubr}. 

Among our 197 runaway stars with $\vperp > 30 \ \kms$, 3 are EBs, 5 are SB2s, and 3 are both EBs and SB2s. This yields a DES binary runaway frequency of $> 6\% \pm 2\%$, which is a substantial lower limit since there are likely many additional binaries we are unable to identify, not only because of selection effects, but also because of the post-SN, two-step ejection mechanism (Section~\ref{subsec:twostep}). \citet{Mason09} give a runaway binary fraction of 43\% and total field binary fraction of 59\%, based on a comprehensive accounting of the observed field O stars. \citet{Lamb16} also estimate a field OB binarity of 59\% $\pm 12$\% for a small subset of the RIOTS4 survey, and \citet{Chini12} find binary frequencies of $69\% \pm11\%$ and $43\% \pm13\%$ for O-star runaways and field stars, respectively. 

On the other hand, the predicted binary frequency of runaway O-stars is around $\sim$30\% \citep{PeretsSubr, OhKroupa}, and slightly higher for lower-mass primaries. While this is consistent with our lower limit on the DES binary runaway frequency of 6\%, the more comprehensive accounting in the above studies shows values that are roughly double the predicted ones. In any case, given that non-compact runaways must originate from dynamical ejections, the kinematics and frequency of this population are consistent with a dominant population of DES ejections, as found above.

\section{Stellar Mass Analysis} \label{sec:mass}
\subsection{Mass Estimates} \label{subsec:Estimates}

Since the dynamical and SN ejection mechanisms predict contrasting relationships between mass and runaway velocity (Section~\ref{subsec:Predictions}), we obtain spectroscopically-determined masses of the RIOTS4 stars to further evaluate the allocation between the two ejection processes. Effective temperatures (\teff), luminosities ($L$), and stellar masses are calculated following an approach similar to that described in \cite{Lamb13}, but updated using the stellar evolutionary models for rotating stars of \cite{Brott}. In particular, \teff\ are derived according to the spectral types published in \citet{Lamb16} with conversions of spectral type to \teff\ from \cite{Massey05} for O-type stars and from \cite{Crowther} for B-type objects. Bolometric magnitudes ($M_{\rm{bol}}$) and luminosities were estimated following \cite{Massey02}, based on $V$ magnitude and  adopting a distance modulus (DM) of 18.9 \citep{Harries03}. The bolometric correction (BC) was obtained as, BC\,$ = 27.99\,-\,6.9\,\log$\,\teff  \citep{Massey05}. We use the extinction $A_V$ listed in \citet{Lamb13}, extracted from the SMC extinction maps in the Magellanic Clouds Photometric Survey \citep{Zaritsky02}. Stellar masses are estimated from the positions of the stars in the Hertzsprung-Russell (H-R) diagram, compared with the rotating (150\,\kms), single-star, evolutionary tracks of \cite{Brott} for SMC metallicity. The masses are obtained by interpolating between the tracks for the evolved stellar mass values at the observed positions in the H-R diagram. For two stars, M2002-38024 and 59319, we adopt the masses from \citet{Lamb13}, since their positions relative to the \citet{Brott} evolutionary tracks do not permit reliable mass determinations.

\begin{figure}[t]
	\plotone{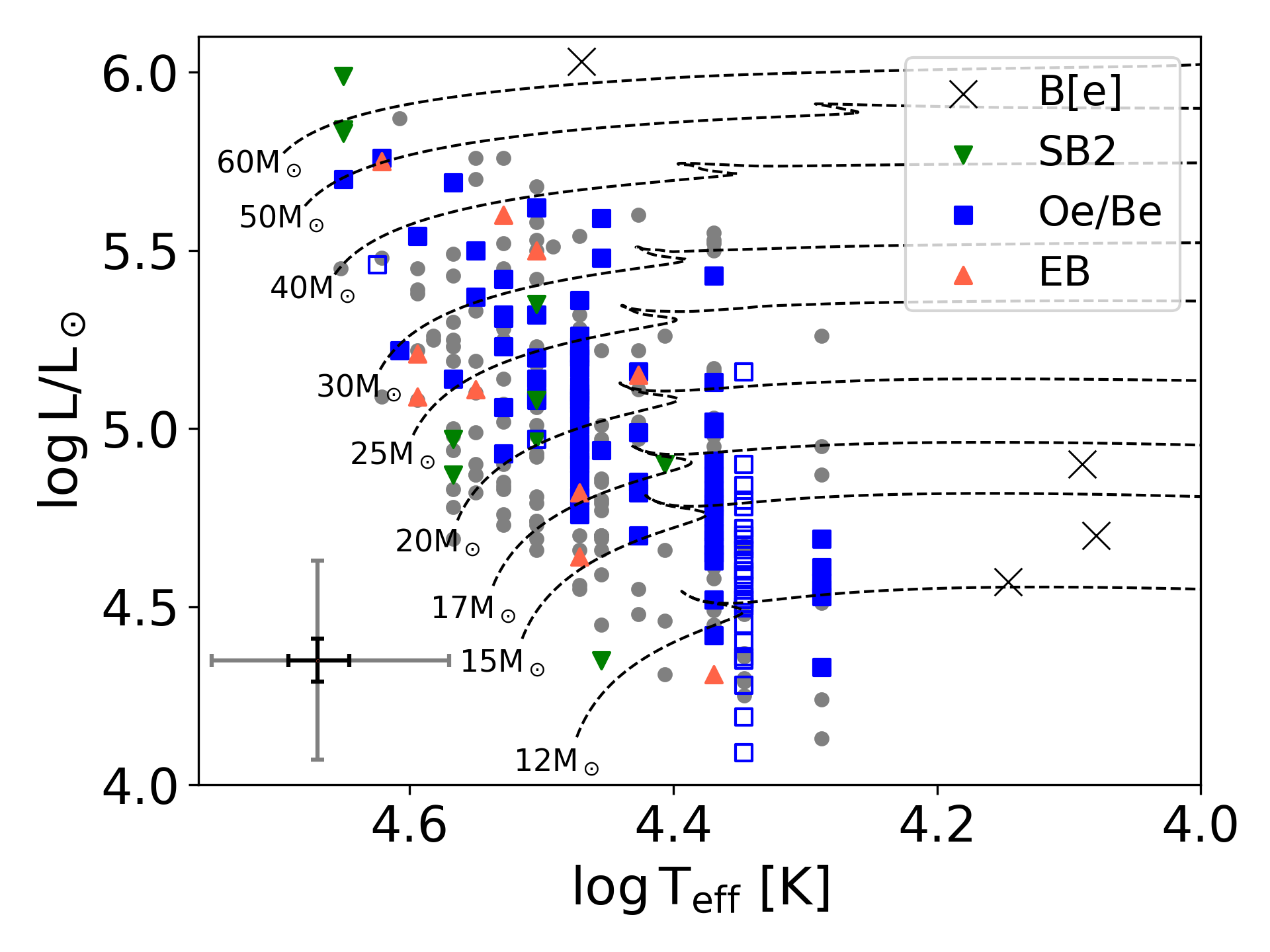}
	\caption{Distribution of the 297 RIOTS4 stars with new mass determinations in the H-R diagram, together with the stellar evolutionary tracks for rotating (150\,\kms) stars from \cite{Brott} computed for SMC metallicity (black dashed lines). EBs, SB2s, B[e] stars, and Oe/Be stars are shown according to the legend. Open squares denote the Oe/Be stars with uncertain mass estimates (see Section~\ref{subsec:Estimates}). Representative error bars are shown for high \teff\ (black, $\log(49.0/\rm kK \pm 2.6/kK)$) and low \teff\ (gray, $\log(20.0/\rm kK \pm 4.3/kK)$). \label{fig:HRdia}}
\end{figure}

Table~\ref{tab:data} gives our mass estimates. The median ratio of our revised masses to those determined by \citet{Lamb13} is $1.03$ for the 107 stars in common, with standard deviation of $0.17$. The greatest uncertainty in determining the stellar masses are the spectral classifications. Of our 297 stars for which we obtained masses, 238 have a well-determined spectral type and thus a well-constrained mass, while 59 do not. For these 59 stars, we adopt the average of the lower and upper mass limits obtained from the limits in the star's range of possible spectral type. The uncertain masses for these stars are flagged with a ``:'' in Table~\ref{tab:data}; these are mostly Be stars that lack a spectral type and SB2s, and also a few Oe stars and ``O + B'' binaries. For the Be stars without a well-constrained spectral type, we calculated the average \teff\ from adopting B0e and B2e spectral types. For SB2s, the masses are calculated for the hotter, more massive star; in the case of "twins", the luminosity is reduced by a factor of two and the quoted mass value is the average of these two. Masses for EBs not identified as SB2s are likely slightly overestimated since these often may be twins. For the four B[e] stars, we obtain spectral types from \citet{Graus12}. Errors for the 238 reliable spectral types are obtained by assuming half a spectral type on either side of that observed, and these errors may therefore be somewhat overestimated. Figure~\ref{fig:HRdia} displays the position of the stars in the H-R diagram together with the \cite{Brott} evolutionary tracks. 

We caution that in what follows, our analyses are based on masses derived from modeled stellar evolution, and not dynamical mass determinations. Moreover, stars in binary systems will have more uncertain masses, as noted in Table~\ref{tab:data}, and if they have experienced binary mass transfer, their positions on the H-R diagram may not provide a meaningful mass estimate \citep[e.g.,][]{Wang2020}. This is especially the case if an active accretion disk is present. In particular, we note that the well-known HMXB SMC X-1 has a well-determined dynamical mass of $15.35 \pm 1.53\ M_{\odot}$ for the primary star, based on accurate X-ray eclipse data \citep{Rawls2011}; whereas our mass obtained from the H-R diagram is twice this value, $32.2 \pm 4.5\ M_{\odot}$. This may be due to the accretion disk amplifying the observed \teff\ and/or luminosity. Furthermore, the star is likely to be rejuvenated by binary mass transfer before the SN event \citep[e.g.,][]{Vinciguerra2020}, altering its stellar structure. Thus, we caution that the masses we obtained for the rest of our HMXBs might also be significantly overestimated.

A power-law fit to the stellar mass distribution for $m\ >\ 20\ M_{\odot}$ gives a slope of $-2.96\ \pm 0.34$, which agrees with the present-day mass function (PDMF) slope of $-2.8$ to $-3.1$ for $m\ >\ 20\ M_{\odot}$ obtained for the subset of RIOTS4 stars studied by \citet{Lamb13}.

\subsection{Stellar Masses and Kinematics} \label{subsec:Mass-Fast}

\begin{figure}[t]
\plotone{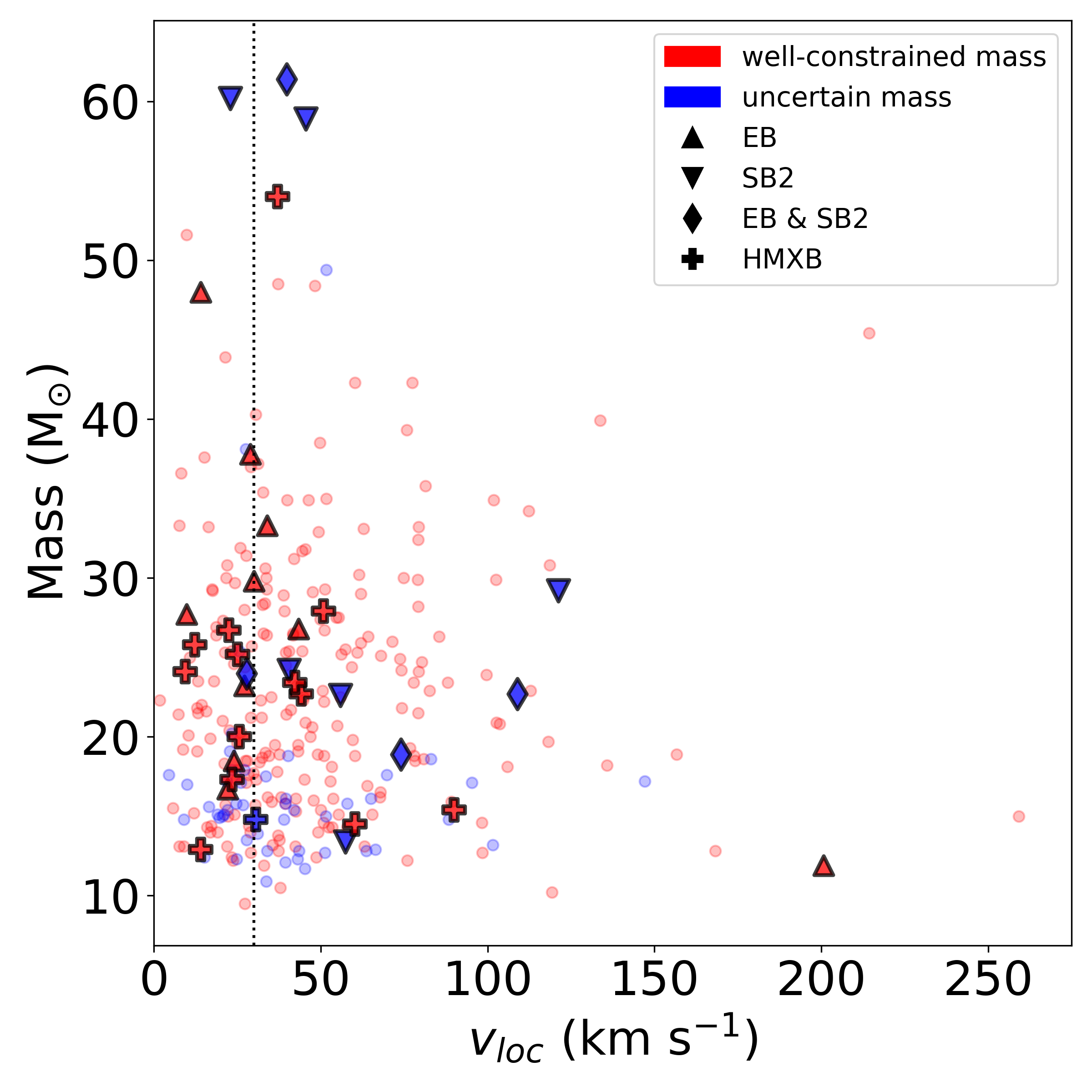}
\caption{Mass versus $\vperp$ for the 299 RIOTS4 stars with mass determinations. Binaries are shown according to the legend. Stars with well-constrained masses are shown in red, while those with uncertain masses are shown in blue. The vertical dotted line depicts our runaway velocity threshold of 30 $\kms$, which we note is near the {\it Gaia} PM detection limit for our sample (27 $\kms$). \label{fig:masspm}}
\end{figure}

\begin{figure}[t]
\plotone{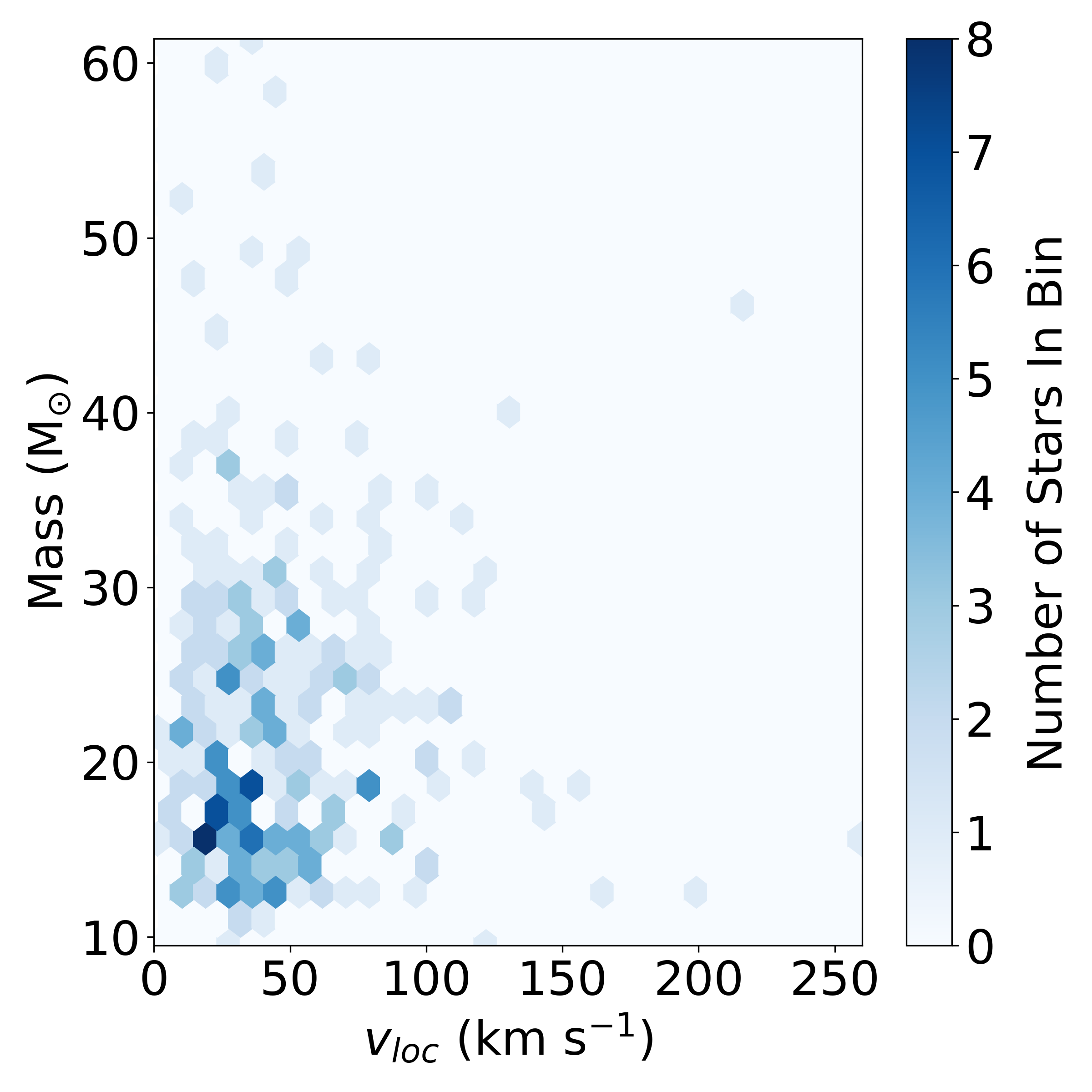}
\caption{Hexbin density plot showing stellar mass and local residual transverse velocity $\vperp$ for the 299 RIOTS4 stars with mass determinations. The distribution suggests a population at low mass and low $\vperp$, and another that is more broadly distributed. \label{fig:hexbin}}
\end{figure}

Our stellar masses are plotted against $\vperp$ in Figure~\ref{fig:masspm}, with the HMXBs, EBs, and SB2s identified. Four stars are both EBs and SB2s. As described in Section~\ref{subsec:Predictions}, the BSS runaway and walkaway velocities are expected to decrease with mass \citep{Renzo19}; while for the DES, the ejection velocity and runaway fraction of stars both increase with mass \citep{Banerjee, PeretsSubr, OhKroupa}. Therefore, we might expect: (1) a slow-moving population, with a range of masses, but predominantly at low masses, resulting from the SN ejection scenario, and (2) a higher-mass, high-velocity population resulting from the dynamical ejection scenario. The density map in Figure~\ref{fig:hexbin} suggests the existence of two such populations, although they are clearly intermixed. We note that the slow population appears to be centered more densely near (30 $\kms$, 18 $M_{\odot}$); this corresponds roughly to our detection and completeness limits, while a second population is also suggested at higher mass and velocity.

\begin{figure}[t]
\plotone{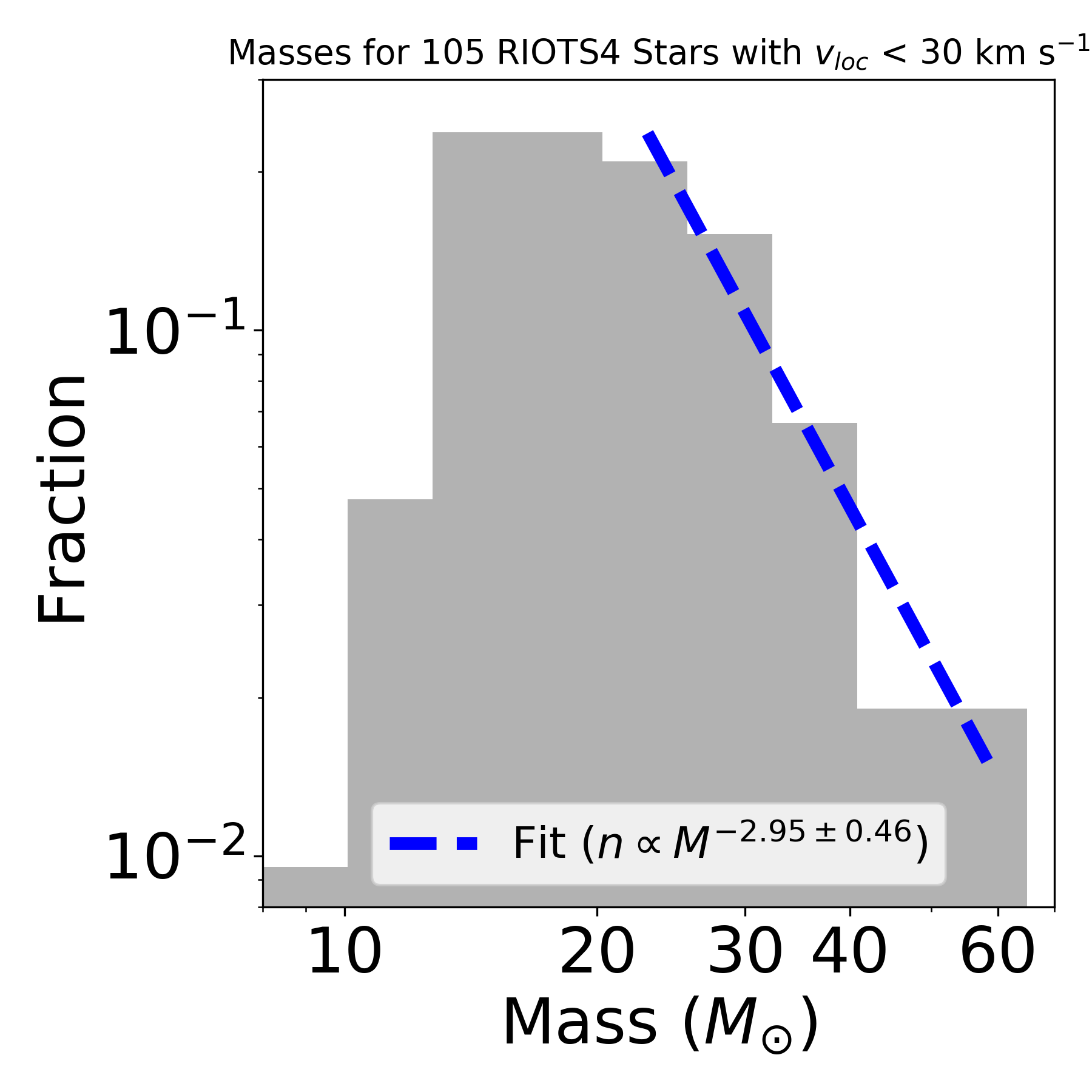}
\plotone{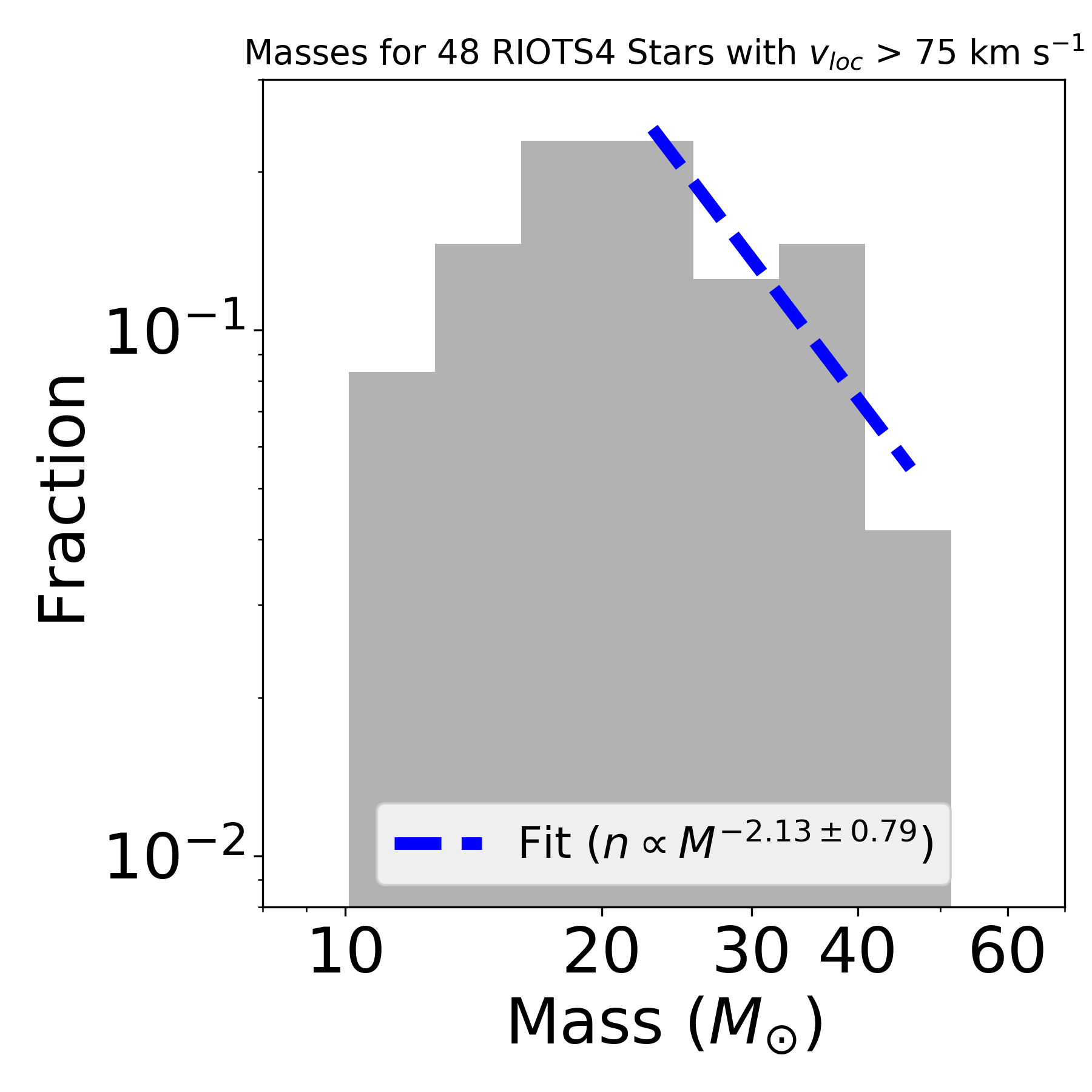}
\caption{Mass distributions for the 105 RIOTS4 stars with $\vperp < 30\ \kms$ and the 48 RIOTS4 stars with $\vperp > 75\ \kms$. Power-law fits are shown for $m\ >\ 20\ M_{\odot}$. \label{fig:masshist}}
\end{figure}

The mass functions for the non-runaway population ($\vperp < 30\ \kms$) and the fast runaways ($\vperp > 75\ \kms$) are shown in Figure~\ref{fig:masshist}. Power-law fits to each stellar mass distribution for $m\ >\ 20\ M_{\odot}$ are shown; the slope for the non-runaway population is  $-2.95\ \pm\ 0.46$, and the slope for the fast runaways is $-2.13\ \pm\ 0.79$. We note that the slope for the non-runaway population is essentially identical to that of the total mass distribution for $m\ >\ 20\ M_{\odot}$ (see above), while the slope for the fast runaways is considerably flatter. Although the error is large, this further supports our understanding that the fastest runaways are produced via the DES mechanism, which favors the ejection of more massive stars, resulting in a flatter mass distribution (Section~\ref{subsec:Predictions}).

For the DES, \citet{FujiiPZ} predict that the median velocity for ejected stars more massive than $\sim 20\ M_{\odot}$ should be relatively constant, a result also found by \citet{Banerjee}, and one that contrasts with expectations for the BSS. Considering only our stars with well-constrained masses, for the mass ranges 18 -- 25 $\msun$ (79 stars), 25 -- 30 $\msun$ (47 stars), and $> 30\ \msun$ (41 stars), we obtain median velocities of 43, 40, and 42 $\kms$, respectively. Thus, the constant median velocity is again consistent with DES dominating our runaway population. We note that these median velocities are somewhat greater than those predicted by \citet{FujiiPZ}, which when projected to 2-D are on the order of 34 $\kms$. This difference may be caused by our sample selection bias against walkaways. However, given the expected significant contamination of slower, BSS walkaways, it may also point to dominant origins in more massive clusters \citep[e.g.,][]{OhKroupa15} or differences in, e.g., the primordial binary parameters assumed by \citet{FujiiPZ} relative to those in the SMC. 

It is interesting to compare our runaway population in Figure~\ref{fig:masspm} with the models of \citet{OhKroupa}. Their Figure~A1 shows that the high-mass, high-velocity parameter space is sensitive to a variety of parameters. At face value, our data are more consistent with binary mass pairing corresponding to either a uniform mass-ratio distribution or ordered pairing. Since dynamical processing in the cluster core likely disrupts most wide binaries, the DES preferentially produces runaway OB stars with close companions, which follow a uniform mass-ratio distribution \citep{Sana12}; whereas wide visual companions are weighted toward smaller mass ratios \citep{Moe17}. This scenario provides enough massive binaries with fairly uniform mass ratios to generate the runaways with velocities on the order of 100 -- 200 $\kms$. We also see non-compact binary runaways at both high mass and high velocity, which are also consistent with these parameters. However, we caution that the models are based on simulations of single-mass clusters, whereas the observed SMC runaways are products from many clusters of different masses, which will affect the velocity distribution \citep{OhKroupa15}.

Although, as discussed above, the runaways are dominated by the DES mechanism, we see in Figure~\ref{fig:masspm} that 7 of the 14 HMXBs have $\vperp > 30\ \kms$. HMXBs are direct tracers of the SN mechanism, and in particular, systems that remain bound, which have faster velocities than single BSS stars. Those with OB secondaries are predicted to have the highest BSS velocities, which can reach speeds of 80 $\kms$ or more \citep{Brandt95, Renzo19}. The median velocity $\vperp$ of our 14 HMXBs is 28 $\kms$, which projected to 3-D is 34 $\kms$, consistent with \citet{Coe05}, who finds an average space velocity of SMC HMXBs $\gtrsim 30\ \kms$. This is greater than the predicted median systemic velocity for NS+MS binaries of only 20 $\kms$ found by \citet{Renzo19}. It is likely due in part to a selection effect favoring the fastest HMXBs, since the fastest runaways are the tightest \citep{Brandt95}, and therefore, most luminous, HMXBs. There are likely many undetected, bound BSS runaways and walkaways.

However, as we shall show in Sections~\ref{subsec:twostep} and \ref{subsec:DESvsBSS}, it is likely that the two-step ejection mechanism is responsible for a significant number of post-SN walkaways and runaways, including HMXBs.  The expected ratio of BSS walkaways to runaways from the models of \citet{Renzo19} is several times higher than our observed ratio of unity. Although our sample is biased against walkaways, the difficulty in generating runaway velocities by the BSS mechanism alone suggests that the two-step mechanism is important in producing our observed BSS runaway population.

\begin{deluxetable*}{ccccccccccccc}
\tablecaption{Kinematic Data and Fundamental Parameters for RIOTS4 Field OB Stars \tablenotemark{a} \label{tab:data}}
\tablehead{\colhead{ID \tablenotemark{b}} & \colhead{SpType \tablenotemark{c}} & \colhead{Subgroup \tablenotemark{d}} & \colhead{$\vperp$ \tablenotemark{e}} & \colhead{err} & \colhead{$M$ \tablenotemark{f}} & \colhead{err \tablenotemark{g}} & \colhead{\vsini} & \colhead{err} & \colhead{\teff} & \colhead{err} & \colhead{$\log\ L$} & \colhead{err} \\ 
\colhead{} & \colhead{} & \colhead{} & \colhead{(\kms)} & \colhead{(\kms)} & \colhead{($M_{\odot}$)} & \colhead{($M_{\odot}$)} & \colhead{(\kms)} & \colhead{(\kms)} & \colhead{(kK)} & \colhead{(kK)} & \colhead{$\log$($L/L_\odot$)} & \colhead{$\log$($L/L_\odot$)}} 
\startdata
107 & Be$_3$ & -,-,-,e & 20 & 21 & 14.9: & 6.3 & \nodata & \nodata & 22.2 & 7.4 & 4.52 & 0.42 \\
1037 & B0.5 V & -,-,-,- & 99 & 30 & 14.6 & 2.5 & 92 & 35 & 26.7 & 3.1 & 4.55 & 0.14 \\
1600 & O8.5 V & E,-,-,- & 43 & 30 & 26.8 & 1.5 & 91 & 17 & 35.5 & 1.5 & 5.11 & 0.05 \\
1631 & B1e$_2$ & -,-,-,e & 51 & 23 & 14.6 & 3.2 & 197 & 11 & 23.4 & 3.7 & 4.71 & 0.19 \\
1830 & B0.5 III & -,-,-,- & 32 & 26 & 22.3 & 4.4 & 86 & 12 & 26.7 & 3.1 & 5.11 & 0.14 \\
2034 & B & -,-,-,- & 21 & 26 & 17.2: & 7.4 & \nodata & \nodata & 22.2 & 7.4 & 4.67 & 0.42 \\
2093 & B1e$_{3+}$ & -,-,-,e & 63 & 27 & 13.1 & 2.8 & \nodata & \nodata & 23.4 & 3.7 & 4.52 & 0.19 \\
3224 & B1e$_{2+}$ & -,-,-,e & 9 & 21 & 19.2 & 4.3 & \nodata & \nodata & 23.4 & 3.7 & 5.0 & 0.19 \\
3459 & O9.5 I & -,-,-,- & 7 & 27 & 33.3 & 2.5 & 238 & 17 & 31.9 & 2.1 & 5.5 & 0.08 \\
3815 & Be$_2$ & -,-,-,e & 147 & 29 & 17.2: & 7.4 & \nodata & \nodata & 22.2 & 7.4 & 4.67 & 0.42 \\
\enddata
\tablenotetext{a}{Table~\ref{tab:data} is available in its entirety in machine-readable format.}
\tablenotetext{b}{From \citet{Massey02}.}
\tablenotetext{c}{Spectral types are from \citet{Lamb16}; except for the four B[e] stars (M2002-29267, 46398, 62661, 83480), which are taken from \citet{Graus12}.}
\tablenotetext{d}{``E'', ``S'', and ``X'' indicate EB, SB2, and HMXB, respectively; ``e'' indicates emission-line star (Oe or Be).}
\tablenotetext{e}{Local residual transverse velocity from Paper~I.}
\tablenotetext{f}{Masses that are uncertain are flagged with ``:'' (see Section~\ref{subsec:Estimates}). Masses flagged with ``n'' are one of the following: The mass for SMC X-1 (M2002-77458) is taken from \citet{Rawls2011}, and those for M2002-38024 and 59319 are taken from \citet{Lamb13}. For the mass of the ``mid'' Oe star M2002-73795 we calculated the average \teff\ from adopting O5e and O7e spectral types, and for the Oe star M2002-75689 we adopted O3e and O9e spectral types. For Be stars without constrained spectral types, we calculate the average \teff\ for B0e and B2e types. For the ``O + B'' binaries (M2002-11238, 22178, 66160) we calculated the average \teff\ from adopting O3 and O9 spectral types, in estimating masses.}
\tablenotetext{g}{Errors are based on $\pm$ half a subtype.  Errors for stars flagged with ``:'' are computed from the upper/lower limits on the spectral type range (see Section~\ref{subsec:Estimates}).}
\end{deluxetable*}

\begin{figure*}[t]
\plotone{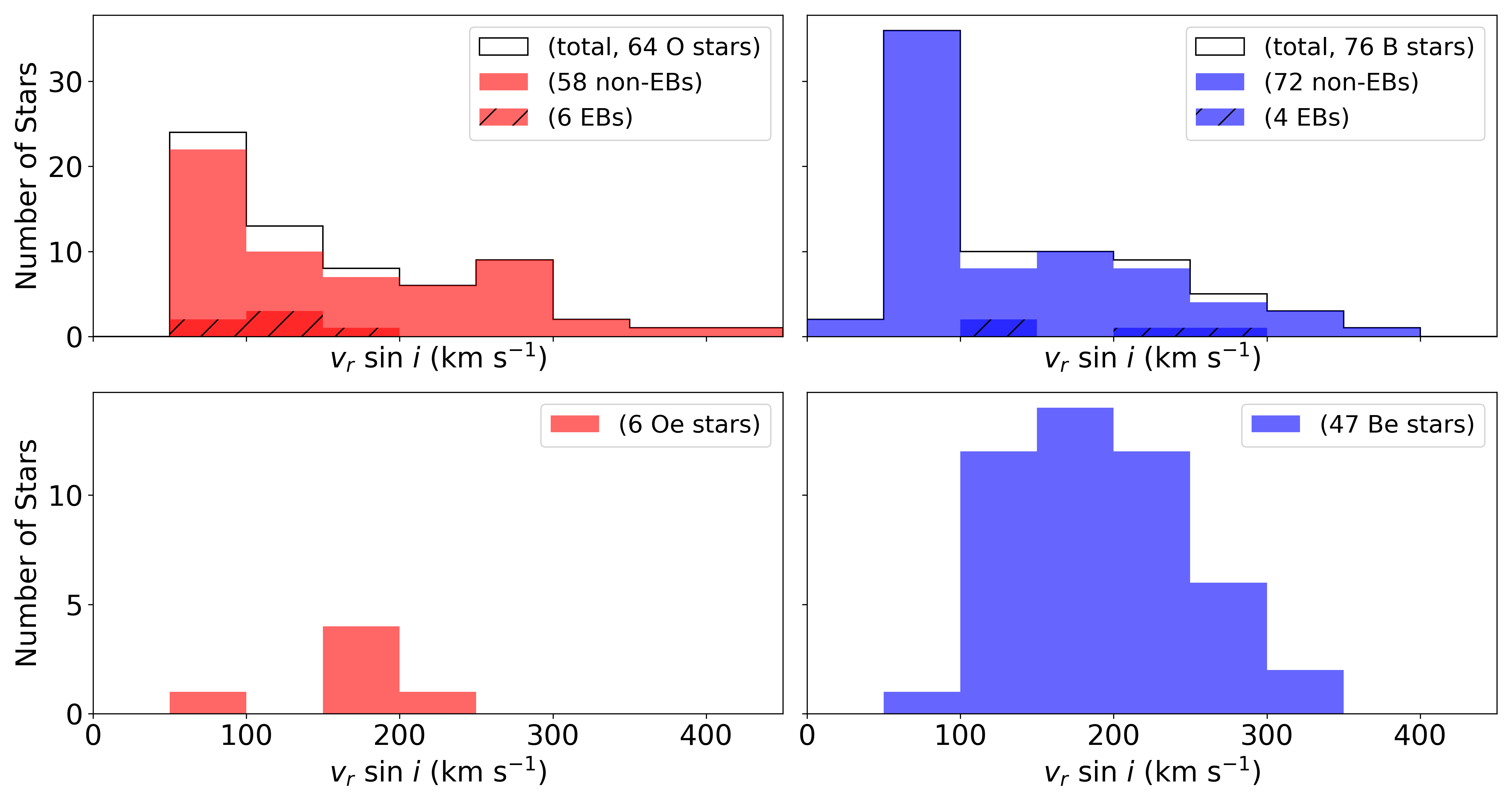}
\caption{Projected rotational velocity distributions for 140 non-Oe/Be stars (top) and 53 Oe/Be stars (bottom). The 8 SB2s with \vsini\ are excluded since their measurements are uncertain. \label{f_vsini}}
\end{figure*}

\begin{figure*}[t]
\plotone{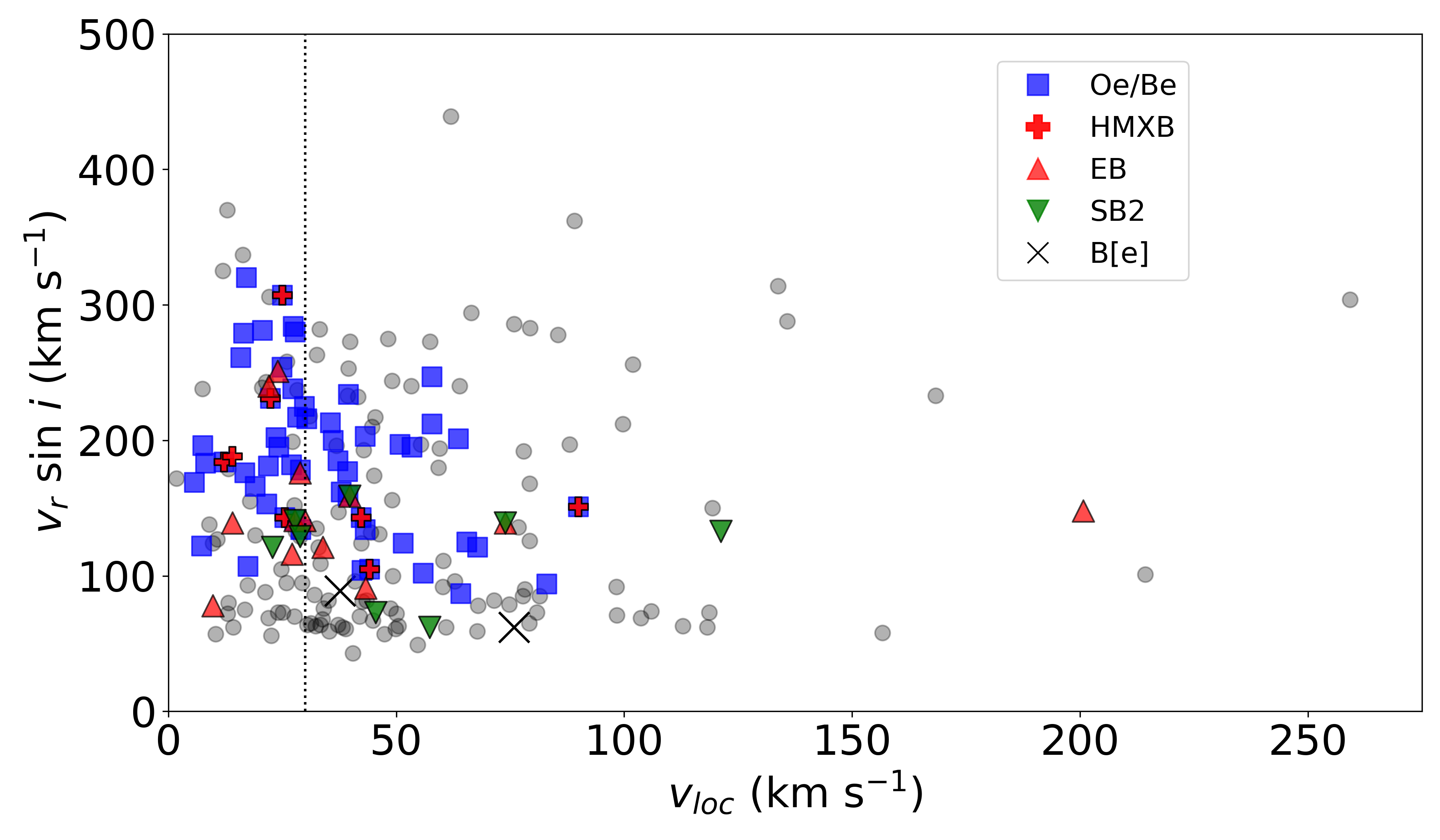}
\caption{Projected rotational velocity \vsini\ versus local residual transverse velocity $\vperp$ for the 201 RIOTS4 stars with \vsini\ measurements. The \vsini\ values for SB2s are upper limits. The vertical dotted line depicts our runaway velocity threshold of 30 $\kms$. \label{f_vsinipm}}
\end{figure*}

\section{Stellar Rotation Analysis} \label{sec:rotation}

We measure projected rotational velocity, \vsini\,, from the medium- to high-resolution spectroscopic data of our sample stars. Most data were taken at $R=3700$ using the IMACS spectrograph at Magellan Observatory, while $\sim$10\% of the sample was observed at much higher resolution, $R=20000$, with the MIKE echelle spectrograph, also at Magellan \citep{Lamb16}. To obtain \vsini, we use the {\sc iacob\_broad} software \citep{Diaz14}, which employs Fourier analysis to differentiate line broadening due to rotation vs macroturbulence, thereby obtaining \vsini\ from individual absorption lines. We consider 13 He I, He II, Si III, and Si IV lines. These lines suffer less from Stark broadening than hydrogen lines, and they are mostly strong features found in the range of spectral types for our sample. We relied especially on the He I lines $\lambda\lambda$4143, 4387, 4471, and 4921. Each star's reported \vsini\ is the median of the measured values from all its available usable lines, weighted by equivalent width, and eliminating outliers. The \vsini\ values are calculated from an average of three lines per star, and we discard stars with \vsini\ measurements of only one suitable line. Our measured \vsini\ for these remaining 201 sample stars are given in Table~\ref{tab:data}. Of the 201 RIOTS4 stars with \vsini\ measurements, 175 were obtained using the IMACS spectrograph, and 26 were obtained using the MIKE echelle spectrograph. The spectral resolution is $\sim$80 $\kms$ for the data at $R=3700$; while the $R=20000$ data have a resolution of $\sim$30 $\kms$.

\subsection{Distribution of \vsini} \label{subsec:vsiniDistrib}

The distribution of \vsini\ for OB stars is known to be bimodal, especially for B stars \citep[e.g.,][]{Wolff07, Braganca12, Dufton13}, and we therefore expect to find such bimodality in our sample. We also expect higher average rotation speeds than typically found in the Milky Way, due to the low metallicity of the SMC, which suppresses a star's ability to lose angular momentum through stellar winds and therefore drives up rotational velocities \citep[e.g.,][]{Maeder00}. The high expected rotation speeds and low metallicity also lead to a high frequency of classical Oe and Be stars, which spin much faster than non-Oe/Be stars \citep[e.g.,][]{Rivinius13, Golden-Marx16}. Oe/Be stars comprise $40\% \pm 4\%$ (= 123/304 stars) of our field OB star sample.

The distribution of our measured \vsini\ is shown in Figure~\ref{f_vsini}, for 140 non-Oe/Be stars (top) and 53 Oe/Be stars (bottom). O/Oe stars are shown in the left panels and B/Be stars on the right. Separating the 47 Be stars from the normal B stars, the bimodality in the \vsini\ distribution apparently corresponds to the different \vsini\ distributions for Be vs non-Be stars. Non-Be stars show a strong peak at low \vsini\ with decreasing numbers at higher values. In contrast, Be stars show a peak at much higher \vsini, consistent with a population dominated by high spin velocities around $250 - 350\ \kms$ that is broadened by \vsini\ projection to lower apparent values. These are likely to be substantial underestimates due to gravity darkening of these highly oblate stars. The O-stars are much fewer, but K-S tests show that the \vsini\ distributions agree statistically with those for the B stars. There are only 6 Oe stars with reliable \vsini\ measurements, due to their tendency to show He I emission, which often causes infill of the features used to measure \vsini\ \citep{Golden-Marx16}.

We can compare our \vsini\ distributions of isolated SMC OB stars to those of OB stars in the SMC Wing published by \citet{Ramachandran2019}. They, too, find that Be stars are much faster rotators than normal OB stars. Our low-velocity peak is situated at our effective resolution limit of $\sim$80 $\kms$, lower than the low-velocity peak of $\sim$120 $\kms$ quoted by \citet{Ramachandran2019}. They also report a mean \vsini\ of 230 $\kms$ for Be stars, which is higher than our value $190\ \kms$. Our fastest-rotating star has \vsini\ of 439 $\kms$, substantially slower than their fastest star at $\sim$550 $\kms$. Thus, in general, their \vsini\ values are slightly higher than ours. \citet{Ramachandran2019} also use the {\sc iacob\_broad} code to determine \vsini, but their sample includes stars of much later spectral types, extending to late B stars. The variation between our results suggests that lower mass B stars may rotate somewhat faster than massive OB stars. This may be consistent with the later spectral types having weaker winds, causing them to retain angular momentum.

\subsection{\vsini\ and Stellar Kinematics} \label{subsec:vsini_runaway}

Stellar rotation provides another important parameter for discriminating between dynamical and SN ejections of OB runaways. As discussed above in Section~\ref{sec:distribution}, the vast majority of BSS ejections are expected to result in slower space velocities, and only 10 -- 20\% of them are predicted to have runaway velocities $\gtrsim 30\ \kms$ at SMC metallicity (Section~\ref{subsec:PredBSS}). The runaway BSS ejections are those originating from the tightest interacting binaries, which have greater orbital velocities at the time of the SN explosion. This implies that mass transfer from the higher-mass primary will spin up the secondary star to rotation speeds near break-up velocity \citep[e.g.,][]{Brandt95, deMink09}. Thus, after the primary explodes, the secondary is ejected with both high space and rotation velocities. This applies to both bound and unbound BSS runaways and is supported by observations \citep{Blaauw93, Hoogerwerf01, Walborn14, MaizApellaniz}. In contrast, single-star dynamical ejections are less likely to be from tight binaries, and hence their rotation speeds should be similar to those of non-runaway stars.

Figure~\ref{f_vsinipm} shows \vsini\ as a function of $\vperp$ for the 201 RIOTS4 stars that have measurements of both quantities. There is a prominent population of stars with low \vsini, near the IMACS resolution limit of 80 $\kms$, that have runaway velocities extending to high values.  This population is strongly inconsistent with the SN ejection mechanism and must correspond to runaways accelerated by the DES. There are 52 stars with \vsini\ $\leq 100\ \kms$ out of the 128 runaway stars with $\vperp \geq\ 30\ \kms$, which sets a lower limit of 41\% on the contribution of DES runaways, consistent with our inference that these strongly dominate over the contribution of BSS runaways (Section~\ref{sec:distribution}; Paper~I).

\subsection{Oe/Be Stars as Post-SN Secondaries} \label{subsec:OeBeRunaways}

We can also use the Oe/Be stars to probe the role of SN ejections in generating runaways. Since Oe/Be stars are expected to be near break-up $v_r$, it seems likely that runaway Oe/Be stars have experienced mass transfer, and therefore many, if not most, acquired their high $\vperp$ from SN kicks \citep[e.g.,][]{deMink13, Boubert18}. In Figure~\ref{f_vsinipm}, Oe/Be stars are indicated with squares and non-Oe/Be stars with circles. Figure~\ref{f_vsinipm} shows that while most Oe/Be stars have $\vperp < 75\ \kms$, many of them are in the runaway regime $> 30\ \kms$. On the other hand, all of the extreme runaways with $\vperp > 90\ \kms$ are non-Oe/Be stars, supporting the expectation that only DES can produce the fastest runaways, and that these tend to be single. The connection between Oe/Be and the BSS is further supported by the fact that very few of our non-compact binaries, which are tracers of the DES mechanism, are Oe/Be stars: only one of our 15 EBs and two of our 11 SB2s are Oe/Be stars, while none of our runaway EBs and only one of our runaway SB2s are Oe/Be stars.

The tendency for fast rotators to be runaways is further shown by the fact that 33 out of our 57 stars with \vsini\ $> 200\ \kms$ are runaways, or $58\% \pm 13\%$. Using their Galactic OB star sample, \citet{MaizApellaniz} found that $13\% \pm 4\%$ of their stars with \vsini\ $> 200\ \kms$ are runaways, using a similar definition of runaways based on velocity dispersion. Since their sample includes both field and non-field stars, the large difference between these two fractions tells us that OB stars with \vsini\ $> 200\ \kms$ are much more likely to be runaways in the field than non-runaways in clusters. This is consistent with our conclusion that most stars with large \vsini\ achieved such high rotation rates due to mass transfer in close binaries, which are more likely to produce runaways. However, we caution that only a subsample of our stars have measured \vsini\ available. Also, high \vsini\ is associated with low metallicity, as discussed in Section~\ref{subsec:vsiniDistrib}, and so a large fraction of DES runaways in our sample will also have high \vsini.

Thus, we may crudely expect that most Oe/Be stars are post-SN runaways \citep[e.g.,][]{McSwain05, deMink13, Boubert18}. On the one hand, the frequency of Oe/Be stars underestimates that of the BSS runaways because not all SN ejections end up as Oe/Be stars, and moreover, the Oe/Be phenomenon likely has a finite lifetime, even for those BSS objects that become Oe/Be stars. On the other hand, runaway DES binaries are also most likely to be tight binaries that undergo mass transfer, possibly producing Oe/Be stars; and single Oe/Be stars may also spin up through other mechanisms, although these should be mostly late-type B stars, which can retain most of their angular momentum due to weaker winds \citep[e.g.,][]{Ekstrom08, deMink09}. The relative magnitude of these effects is not known, but they counteract each other. Thus, it may not be unreasonable to crudely assume that all runaway Oe/Be stars are accelerated by SNe, including two-step ejections, and the remaining runaways are due to dynamical ejections. As noted above, about half of Oe/Be stars have runaway velocities, as we also find for HMXBs (Section~\ref{subsec:Mass-Fast}). This consistency further supports adopting Oe/Be stars as tracers of BSS ejections, and it again reveals an unexpectedly large number of BSS runaways.

There are 69 Oe/Be stars out of the 197 runaway stars in our full sample, therefore implying a BSS runaway fraction of $35\% \pm 5\%$. This crude estimate is larger than our $\sim 20$\% estimate obtained in Section~\ref{sec:distribution}. It does not account for objects in common between DES and BSS, but suggests a slightly larger frequency (see Section~\ref{subsec:DESvsBSS}).

\section{Discussion} \label{sec:discussion}

The observed kinematics of the RIOTS4 runaway stars, and their relation to multiple different parameters including stellar mass, binarity, and \vsini, paint a consistent picture that the SMC field OB runaways are allocated as roughly 70:30 to 80:20 for DES:BSS ejections, respectively. This is based on independent predictions for the products of both of these mechanisms. 

Table~\ref{tab:frequencies} presents our estimates for the frequencies of runaway and walkaway stars in the total SMC OB population that we will show below to be somewhat consistent with both predictions and our observations, assuming a negligible contribution from objects that formed {\it in situ} (Section~\ref{subsec:insitu}). The values in the table represent sub-population frequencies that are determined self-consistently for a single parent population.

The contributions of the various sub-populations are driven primarily by the branching between DES runaways and walkaways, for which we adopt a ratio of 30:70, respectively.  This corresponds to about the maximum allocation for runaways seen in the models of \citet{OhKroupa}. The frequencies are also affected by the total fraction of ejected stars (Section~\ref{subsec:incompleteness}) and the DES binary ejection fraction, which we set to 30\%, again guided by models \citep[e.g.,][]{PeretsSubr, OhKroupa}.

As noted earlier, BSS ejections have lower frequencies. In Section~\ref{subsec:PredBSS}, we estimated the steady-state BSS OB runaway frequency to be 0.020 (Table~\ref{tab:frequencies}, BSS All Ejected); for BSS ``walkaways,'' the corresponding frequency estimate is 0.22. This is based on the model of \citet{Renzo19} for SMC metallicity and accounts for both bound and unbound binary components, which have frequencies of 0.0080 and 0.012, respectively, for this model. However, since that work does not consider clusters, their definition of ``walkaways'' does not correspond to objects ejected from clusters, but rather, simply to post-SN objects that are not runaways.  In this work, we use the term ``walkaways'' to refer to objects that are unbound from clusters but have $\vperp < 30\ \kms$, and thus which populate the field. Their number depends on their velocity distribution and cluster density profiles, which are model-dependent, but we can crudely estimate that about half of the predicted, BSS so-called ``walkaways'' become true walkaways, based on BSS models by \citet{Renzo19} and DES models by \citet{OhKroupa}.  We therefore obtain their rough frequency contribution to be $\sim$0.11 (Table~\ref{tab:frequencies}, BSS All Ejected).  

The BSS runaway and walkaway frequencies are summarized in Table~\ref{tab:frequencies}, which also gives the break-down between objects remaining bound vs unbound by the SN explosions, from the SMC model of \citet{Renzo19}. The rest of the values in Table~\ref{tab:frequencies} are estimated in what follows, and assume that 24\% of the steady-state OB population is post-SN (Section~\ref{subsec:PredBSS}).

\subsection{Walkaway Incompleteness} \label{subsec:incompleteness}

In Section~\ref{sec:distribution}, we found that about 70\% of our field OB stars are runaways, implying an overall frequency of 0.18 for a field OB-star fraction of about 25\% (Table~\ref{tab:frequencies}, Total Observed).  We therefore take walkaways to be 30\% of the field, but as noted in Section~\ref{sec:distribution}, we have a large selection bias against walkaways, requiring a correction factor of $\sim$2.4. This therefore implies that our observationally derived walkaway frequency is also 0.18, or roughly equal to that of the runaways (Table~\ref{tab:frequencies}, Total Observed). This uniform ratio of walkaways to runaways is harder to match with predictions, since both DES and BSS predict many more walkaways than runaways. One possible issue could be that the observed ratio is affected by spurious runaways due to poorly quantified effects with {\it Gaia} proper motions.  

It is also plausible that there is further incompleteness among the walkaways, in addition to the selection bias. An additional observational bias against both walkaways and cluster stars is quite likely in the \citet{Oey04} survey from which RIOTS4 is drawn, due to extinction and crowding near clusters \citep[e.g.,][]{Schoettler20}. We note that the \citet{Oey04} survey has a total of 1360 OB candidates, whereas in contrast, the survey by \citet{Evans04} identified about 2500 OB stars, including 23 O stars that were fainter than the RIOTS4 selection criterion. A large fraction of their OB stars have uncertain spectral types and are unlikely to be in the RIOTS4 spectral range, but the number of OB candidates is consistent with RIOTS4 having a significant incompleteness for stars in, and near, clusters.  

This would also alleviate the potentially high fraction of field stars in our analysis.  Our corrected observed total frequency of ejected objects is $\sim$36\% of the entire OB population, while observations suggest field star populations closer to 20 -- 30\%. On the other hand, the high observed ejected binary frequency (Section~\ref{subsec:binary}) supports a high ejection fraction. The reported field fractions depend on selection criteria; the youngest unbound objects, especially walkaways, are usually difficult to identify when still spatially within the cluster's projected area.  

Thus, the discrepancy between the predicted and observed ratio of walkaways to runaways is mitigated by observational effects, but it does remain notable.  We return to this issue below in Section~\ref{subsec:DESvsBSS}.

We adopt a DES frequency of 0.33 for OB stars ejected from clusters, which yields the total observed ejection frequency of 0.36 when accounting for BSS ejections (Section~\ref{subsec:DESvsBSS}). The DES runaway:walkaway branching of 30:70 adopted above yields total DES runaway and walkaway frequencies of 0.10 and 0.23 (Table~\ref{tab:frequencies}, DES All Ejected), respectively. Table~\ref{tab:frequencies} presents the resulting estimates for the frequencies of DES runaway and walkaway, pre- and post-SN sub-populations and binaries in the SMC, assuming the post-SN and DES binary frequencies adopted above.

Some of the 0.67 of OB stars that are not dynamically ejected from clusters are ejected by the BSS mechanism.  With expected, respective frequencies of runaway and walkaway BSS ejections of 0.020 and 0.11 noted above, this yields pure BSS ejection frequencies of 0.013 and 0.074 (Table~\ref{tab:frequencies}, Total pure BSS).

\subsection{Two-Step Ejections} \label{subsec:twostep}

Ejected, non-compact binaries are most likely to be the tightest systems, and therefore, these are destined to be progenitors of the ``two-step ejection'' mechanism \citep{Pflamm-Altenburg10}, which re-accelerates the surviving star upon the SN explosion of the primary.  These two-step ejections are therefore a subset of both DES and BSS ejections. We note that out of our total 11 runaway EBs and SB2s, 8 have at least one O-star, increasing the likelihood that the secondary is also massive. The peak ejection age of $\sim$1 Myr for binaries \citep{OhKroupa} is a relatively small fraction of typical OB star lifetimes (3 -- 20 Myr), and so the vast majority of non-compact binary ejected systems will experience their first SN after the ejection event \citep{Pflamm-Altenburg10}.

The two-step process generates much higher space velocities, up to 1.5 -- 2$\times$ faster than can be achieved by the BSS alone \citep{Pflamm-Altenburg10}. Interestingly, our fastest HMXB is the well-known object SMC X-1 (M2002-77458), which has $\vperp = 90\ \pm\ 31\ \kms$ and $m = 15.4\ \pm\ 1.5\ M_{\odot}$ (Figure~\ref{fig:masspm}; Table~\ref{tab:data}). Its proper motion must be confirmed, and we caution that there is inherent uncertainty in determining $\vperp$ relative to other field OB stars. Paper~I obtains a radial velocity for this star of 29 $\kms$ relative to SMC systemic, yielding a total space velocity of 95 $\kms$. The extreme speed relative to the small number of HMXBs is suggestive that the two-step ejection process may have played a role in this object's velocity in particular, although its measurement error remains consistent with an expected velocity around 70 $\kms$ for an object with its parameters \citep[e.g.,][]{Brandt95}. Its short orbital period of 3.9 days \citep{HaberlSturm} is also consistent with a high runaway velocity.

The post-SN binaries accelerated by DES correspond to two-step ejections, and for the parameters in Table~\ref{tab:frequencies}, their walkaway frequency is 0.017 (Table~\ref{tab:frequencies}, DES Post-SN, binaries). Based on the DES and BSS walkaway velocity distributions \citep{OhKroupa, Renzo19}, we roughly estimate that 1/3 of these two-step walkaways are re-accelerated to runaway velocities, which yields a frequency contribution 0.006 of new runaways unaccounted for by either DES or BSS models. Together with the original, runaway post-SN binaries, this gives a total two-step runaway frequency of 0.013 (Table~\ref{tab:frequencies}, DES Two-step).

Two-step ejections may generally be observed as BSS objects. Thus, by adding the contributions from two-step runaways and pure BSS runaways, we estimate the frequency of runaways observed as BSS objects to be 0.026 (Table~\ref{tab:frequencies}, BSS With two-step). This implies that {\it two-step runaways may correspond to at least half of all BSS runaways.} These estimates are model-dependent, but in any case, we see that two-step runaways are likely a substantial fraction of BSS runaways, and could dominate if the binary fraction is significantly larger than the assumed value of 30\%, as appears to be the case (Sections~\ref{subsec:binary} and \ref{subsec:DESvsBSS}). The resulting BSS population frequencies are summarized in Table~\ref{tab:frequencies}. 

\begin{deluxetable}{lcc}
\tablecaption{Runaway and Walkaway Frequencies: \\
Model-driven\tablenotemark{a} \label{tab:frequencies}}
\tablehead{\colhead{} &\colhead{Runaways} & \colhead{Walkaways}} 
\startdata
DES & &  \\
\quad All Ejected & 0.10 & 0.23 \\
\quad {\bf Pre-SN} & {\bf 0.075} & {\bf 0.18} \\
\quad Post-SN & 0.024 & 0.055 \\
\quad Pre-SN, binaries & 0.023 & 0.053 \\
\quad Post-SN, binaries & 0.0071 & 0.017 \\
\quad Two-step\tablenotemark{b} & 0.013 & 0.011 \\
\hline
BSS & &  \\
\quad Unbound & 0.012 & 0.052 \\
\quad Bound & 0.0080 & 0.060 \\
\quad All Ejected & 0.020 & 0.11\tablenotemark{c} \\
\quad Total pure BSS & 0.013 & 0.074 \\
\quad {\bf With two-step} & {\bf 0.026} & {\bf 0.086} \\
\hline
Total Predicted & {\bf 0.10} & {\bf 0.26}  \\
Total Observed & 0.18 & 0.18\tablenotemark{d}  \\
\hline
DES/BSS ratio & {\bf 2.9} & 2.1 \\
Pre-SN binaries, sub-pop\tablenotemark{e} & {\bf 0.22} & {\bf 0.20} \\
\hline
\enddata
\tablenotetext{a}{Estimated frequencies of runaway and walkaway sub-populations in the total SMC OB population, adopting: (1) a DES ejection fraction of 0.33, (2) a DES runaway:walkaway branching ratio of 30:70, and (3) a DES binary frequency of 0.3. BSS frequencies are from \citet{Renzo19} for SMC metallicity. Values compared to observations are boldface.}
\tablenotetext{b}{Assuming that 1/3 of DES binary post-SN walkaways become two-step runaways (Section~\ref{subsec:twostep}).}
\tablenotetext{c}{Assuming that half of BSS non-runaways are ejected from clusters (Section~\ref{subsec:incompleteness}).}
\tablenotetext{d}{Corrected by factor 2.4 for incompleteness due to field sample selection effect (Section~\ref{subsec:incompleteness}).}
\tablenotetext{e}{Non-compact binary frequencies within the respective runaway and walkaway sub-populations.}
\end{deluxetable}

\begin{deluxetable}{lcc}
\tablecaption{Runaway and Walkaway Frequencies: \\
Forced Match to Observations\tablenotemark{a} \label{tab:frequencies2}}
\tablehead{\colhead{} &\colhead{Runaways} & \colhead{Walkaways}} 
\startdata
DES & &  \\
\quad All Ejected & 0.15 & 0.15 \\
\quad {\bf Pre-SN} & {\bf 0.11} & {\bf 0.11} \\
\quad Post-SN & 0.036 & 0.036 \\
\quad Pre-SN, binaries & 0.11 & 0.11 \\
\quad Post-SN, binaries & 0.036 & 0.036 \\
\quad Two-step & 0.048 & 0.024 \\
\hline
BSS & &  \\
\quad Unbound & 0.012 & 0.052 \\
\quad Bound & 0.0080 & 0.060 \\
\quad All Ejected & 0.020 & 0.11 \\
\quad Total pure BSS & 0.014 & 0.078 \\
\quad {\bf With two-step} & {\bf 0.062} & {\bf 0.10} \\
\hline
Total Predicted & {\bf 0.18} & {\bf 0.22}  \\
Total Observed & 0.18 & 0.18  \\
\hline
DES/BSS ratio & {\bf 1.8} & 1.1 \\
Pre-SN binaries, sub-pop & {\bf 0.65} & {\bf 0.53} \\
\hline
\enddata
\tablenotetext{a}{Frequencies calculated as in Table~\ref{tab:frequencies}, but adopting: (1) a DES ejection fraction of 0.30, (2) a DES runaway:walkaway branching ratio of 50:50, and (3) a DES binary frequency of 1.0. Values compared to observations are boldface.}
\end{deluxetable}

\subsection{DES vs BSS Ejections in the SMC} \label{subsec:DESvsBSS}

With the revised values for BSS due to the effects of the two-step mechanism, Table~\ref{tab:frequencies} shows that the total values for predicted runaways and walkaways are only slightly adjusted to 0.10 and 0.26, respectively, from the original DES ejection frequencies (Table~\ref{tab:frequencies}, Total Predicted). The total predicted frequencies, as well as the DES/BSS ratios, are calculated using the frequencies of pre-SN DES objects and BSS objects including two-step ejections, since these correspond to what is observed. The resulting walkaway-to-runaway ratio is $\sim$2.6, whereas the observed ratio, corrected for walkaway incompleteness due to selection bias, is $\sim$1. We see that the total walkaway-to-runaway ratio is strongly dominated by the DES mechanism, and our adopted branching ratio of 70\% walkaways vs 30\% runaways allows about the maximum value for runaways that is plausible from the models of \citet{OhKroupa}. Although all models predict several times more walkaways than runaways, their ratio is difficult to decrease, a problem that is exacerbated by the cluster mass function \citep{OhKroupa15, OhKroupa}. If anything, the observed ratio is too large since our 2-D velocity threshold defining runaways is more stringent than the 3-D threshold used in the models. As discussed in Section~\ref{subsec:incompleteness}, accounting for possible spurious runaways and additional incompleteness due to extinction and crowding may help to resolve the discrepancy. However, our results are suggestive of a DES runaway-to-walkaway production that is higher than expected.

We also see that the total predicted non-compact binary ejection frequency is 0.21, when calculating the weighted average for both runaways and walkaways. This is low compared to the reported observed value of 0.59 for field stars \citep{Mason09,Lamb16}, indicating a significantly higher value for the binary ejection rate than the adopted value of 0.3. Increasing this parameter would decrease the DES-to-BSS runaway ratio, counting only pre-SN objects for DES. It would also decrease the BSS walkaway-to-runaway ratio, but these effects could be counteracted by increasing the DES runaway vs walkaway allocation, as suggested above.  Increasing the total ejection frequency also slightly increases the DES-to-BSS ratio and decreases the walkaway-to-runaway ratio.

We can explore the effect of forcing the parameters to match the observations by adopting, for example, a DES runaway-to-walkaway branching ratio of 50:50 and DES binary fraction of 1.0.  This would imply that all ejected systems are expelled before their SNe or dynamical binary disruptions. We adopt a DES ejection fraction of 0.3, which yields a total ejection fraction of 0.39.  Table~\ref{tab:frequencies2} gives revised estimates for the frequencies of the various sub-populations, calculated in the same way as in Table~\ref{tab:frequencies}. These input values produce results that are more in line with our observations:  the total walkaway-to-runaway ratio has decreased to 1.2, which is much closer to the observationally derived ratio of 1.0 (Table~\ref{tab:frequencies2}); and the total ejected binary fraction is 0.58, which agrees with the observed value of 0.59 \citep{Mason09,Lamb16}. The DES/BSS runaway ratio is 1.8 (Table~\ref{tab:frequencies2}, DES/BSS ratio), which is still consistent with our observations, in particular, the frequency of Oe/Be stars (Sections~\ref{subsec:OeBeRunaways} and \ref{subsec:origin_oebe}).

Thus overall, our analysis suggests a ratio of DES to BSS runaways of $\sim 2-3$. Our data suggest that DES predictions may underestimate runaway production relative to walkaways. However, there could also be underlying issues with incompleteness and other observational biases, as described in Section~\ref{subsec:incompleteness}. We also caution that the DES models of \citet{OhKroupa} and BSS models of \citet{Renzo19} are independent, and there are likely minor inconsistencies beteween them, and additional physical relationships between DES and BSS ejections that are unaccounted for.

Observations of the field binary frequency more strongly suggest that the DES binary ejection fraction is higher than adopted based on the models of \citet{OhKroupa}.  This may be due to the importance of lower-mass clusters, which eject binaries at higher rates \citep{OhKroupa15}. Increasing the DES binary fraction also increases the frequency of two-step ejections, and can do so substantially.  Whereas in Table~\ref{tab:frequencies} the number of two-step and pure BSS runaways are equal, in Table~\ref{tab:frequencies2} two-step ejections are more than 4$\times$ the number of pure BSS runaways.  Thus, if the DES binary ejection fraction is indeed high, then {\it two-step ejections likely dominate BSS runaways.} 

This can explain the unusually large frequency of BSS runaways (Sections~\ref{subsec:Mass-Fast} and \ref{subsec:OeBeRunaways}). From fiducial models, we expect a walkaway-to-runaway ratio of $\sim$6 for pure BSS ejections and $\sim$3 including two-step ejections (Table~\ref{tab:frequencies}), recalling that runaways are difficult to produce via BSS.  But increasing both the DES binary ejection and DES runaway frequencies strongly increases the two-step contribution to the BSS runaway population, bringing the BSS walkaway-to-runaway ratio closer to the observed ratio of $\sim 1$ (Table~\ref{tab:frequencies2}). Allowing for incompleteness in the observed number of walkaways, the contribution of two-step runaways can easily account for the high observed rate of BSS runaways.

\subsection{In-Situ OB Star Formation} \label{subsec:insitu}

The expected number of walkaways relative to our survey data implies that the frequency of any field OB stars that formed \textit{in situ} must be small. We will address this in a forthcoming work that identifies such objects for our RIOTS4 targets by searching for associated small clusters, and confirms that there are few detections \citep{VargasSalazar20}. These findings support suggestions by \citet{Gvaramadze12} that essentially all field OB stars are ejected systems.

\subsection{The Origin of Oe/Be Stars} \label{subsec:origin_oebe}

The origin of classical Oe/Be stars has long been a puzzle, and is often attributed to non-radial pulsations and/or magnetic phenomena \citep[see, e.g., the review by][]{Rivinius13}. However, a model that historically has received less attention is that Oe/Be stars might simply originate as objects that acquire their high rotation velocities through binary mass transfer \citep[e.g.,][]{Pols91, VanBever97, McSwain05, deMink13}. Modern understanding of massive binary properties and statistics provide new leverage for this model. In particular, binary population synthesis models show that the frequency of Oe/Be runaways in the Milky Way is consistent with all of them having formed through the post-mass-transfer model \citep{Shao14, Boubert18}, and observations are consistent with a prevalence of compact companions \citep{Klement19} and a lack of main-sequence companions \citep{Bodensteiner20}.

We showed in Section~\ref{subsec:OeBeRunaways} that the statistics and kinematics of classical Oe/Be stars in our sample are fully consistent with this population largely corresponding to post-SN secondaries that remain after the original primary has exploded. This is further supported by their statistical similarities to HMXBs. Indeed, all but one of the HMXBs in our sample are Be stars. As noted above, we find that 7 of our 14 total HMXBs, and 69 of our total 123 Oe/Be stars, have runaway velocities $> 30\ \kms$. These numbers yield walkaway to runaway ratios of $1.0$ and $0.9$, respectively, which agree well with each other, further supporting the premise that the Oe/Be stars correspond to BSS ejections. The predicted ratio of BSS walkaways to runaways estimated in Table~\ref{tab:frequencies2} is $\sim$ 1.6, which is larger than the observed value of $\sim 1$, a difference that is easily attributed to walkaway incompleteness, which is not accounted for. Furthermore, the predicted ratio of DES to BSS runaways in Table~\ref{tab:frequencies2} is 1.8, which agrees with the observed ratio of 1.9 (128 DES / 69 BSS), assuming all our 69 runaway Oe/Be stars are from the BSS. For walkaways, the predicted ratio is $\sim$1.1 (Table~\ref{tab:frequencies2}), again in remarkable agreement with the observed ratio of 1.0 (53 DES / 54 BSS). These statistics include the contribution of two-step ejections, which appear to dominate the BSS, and therefore, Oe/Be population.

\citet{Renzo19} note that most bound, post-SN systems have tight orbital periods, and likely undergo mass transfer before the SN event, which thereby provides not only a simple explanation for the extreme rotational velocities, but also a prediction that they must necessarily occur at relatively high frequencies. This suggests that most therefore are spun up to velocities exceeding the critical value, thus generating the excretion disks responsible for the line emission. It also suggests that most of these would also go through a HMXB phase. The consistency of the Oe/Be stars with the BSS statistics suggests that the disks are long-lived. Moreover, this model also provides a simple explanation for the strong bimodality in \vsini\ (Section~\ref{subsec:vsiniDistrib}) that is not explained by other models for the origin of Oe/Be stars \citep{deMink13}. Thus, while the Be phenomenon may also originate through other mechanisms, our data strongly support the post-mass-transfer model, with the vast majority corresponding to surviving, post-SN objects.
\vspace{10mm}

\section{Summary} \label{sec:summary}

One of the most enduring topics in stellar kinematics is the existence of massive runaway stars, O- or B-type stars traveling faster than 30 $\kms$. There are two mechanisms capable of ejecting OB stars from their birth clusters into the field at such velocities: the dynamical ejection scenario (DES) and the binary supernova scenario (BSS). Our work provides a first estimate for the relative contributions from these two ejection mechanisms for a complete sample of field stars in an external galaxy, clarifying the interaction histories of massive stars.  Our analysis is based on our sample of 304 SMC field OB stars from the spatially complete RIOTS4 survey, examining: (1) local residual transverse velocities, $\vperp$, of 304 stars (Section~\ref{sec:distribution}, Figure~\ref{f_PM_distrib}), (2) masses for 299 stars (Section~\ref{subsec:Estimates}, Figures~\ref{fig:masspm} -- \ref{fig:masshist}), and (3) projected rotational velocities, \vsini\,, for 201 stars (Section~\ref{sec:rotation}, Figure~\ref{f_vsinipm}). We obtained spectroscopically determined masses based on stellar evolutionary models for rotating stars of \citet{Brott}, and we measured \vsini\ using {\sc iacob\_broad} software \citep{Diaz14}, which employs Fourier analysis to differentiate line broadening due to rotation from macroturbulence (Table~\ref{tab:data}). 

The distributions of both our masses and \vsini\ are generally consistent with expectations. The mass function we obtain for $m > 20\ M_{\odot}$ yields a slope of $-2.96\ \pm\ 0.34$, which agrees with the RIOTS4 PDMF obtained by \citet{Lamb13}. Our \vsini\ distributions (Figure~\ref{f_vsini}) confirm that non-Be stars peak at low \vsini\ with decreasing numbers at higher values, whereas Be stars peak at much higher \vsini.

To estimate the fraction of DES runaways, we adopt the predicted velocity distribution from \citet{PeretsSubr} and extrapolate from the high-velocity tail, which is dominated by DES ejections. We independently estimate the frequency of BSS runaways, based on binary population synthesis models of \citet{Renzo19}. For SMC metallicity, these predict relative DES:BSS contributions around 80:20.

Non-compact, binary runaways, such as EBs and SB2s, are a direct probe of the DES mechanism. We find characteristics of our non-compact binaries that are consistent with predictions of binaries produced by the DES \citep{PeretsSubr, OhKroupa}: (1) in general, our non-compact binary frequency decreases with increasing velocity; and (2) our highest velocity binaries have speeds commensurate with predictions on the order of a few 100 $\kms$. The fraction of our runaways that are non-compact binaries ($> 6\% \pm 2\%$) is also consistent with predictions, although it is a substantial underestimate. Overall, our non-compact, binary runaway population is consistent with a dominant population of DES ejections, as found above.

This is also supported when the stellar masses are considered with the kinematics. The distribution of mass vs $\vperp$ supports the presence of two populations (Figure~\ref{fig:hexbin}), one corresponding to slow stars at a range of masses, and another with a broader range of velocities and skewed to somewhat higher masses. The fastest runaways show flatter mass functions (Figure~\ref{fig:masshist}), as expected from DES predictions that the highest-mass stars are preferentially ejected; while median velocities remain relatively constant as a function of mass.

Stellar rotation provides another important parameter for discriminating between the products of the DES and the BSS. Runaway OB stars produced by the BSS come from the tightest interacting massive binaries since they have the highest pre-SN orbital velocities, thus leading to high rotational velocities and that can cause excretion of circumstellar material driving the Be phenomenon. The distribution of \vsini\ vs $\vperp$ again shows the two types of runaway stars (Figure~\ref{f_vsinipm}), with one population at low \vsini\ ($\leq 100\ \kms$) showing runaway velocities extending to high $\vperp$ (DES); and another with much higher \vsini\ and somewhat lower transverse velocities (BSS). Runaway Oe/Be stars appear to correspond to BSS systems, and so we can use them to represent the BSS runaway population (Section~\ref{subsec:OeBeRunaways}). This suggests a DES:BSS allocation of $\sim$70:30.

Overall, analysis of the kinematics expected for each population and the statistics of Oe/Be stars imply that \textbf{dynamical ejections dominate, with the ratio of DES to BSS runaways $\sim 2-3$ in the SMC.} A breakdown of the runaway and walkaway populations that is model-driven, but somewhat consistent with our observations is given in Table~\ref{tab:frequencies}, and another that forces a match to the observations is given in Table~\ref{tab:frequencies2}. Our results suggest that the DES runaway production rate relative to walkaways may be higher than predicted, although incompleteness in our walkaway population may alleviate the former. Comparing these results to observations of the field binary frequency in the literature also suggests that the DES binary ejection rate is high for ejected systems.

Two-step ejections \citep{Pflamm-Altenburg10} are a subset of both DES and BSS populations, and {\bf they are a substantial fraction, on the order of half, or more, of all BSS runaways.}  Moreover, on the order of 1/4 of BSS runaways may be objects accelerated above the runaway threshold by the two-step process, which are therefore new runaways not accounted for by either DES or BSS models. Two-step runaways are fundamentally linked to the DES binary fraction, which observations imply is large.  This suggests that {\it two-step runaways may substantially dominate the BSS runaway population.}

The large number of expected walkaways in our sample also implies that any contribution of field OB stars that formed \textit{in situ} is small. This is consistent with results from our search for such objects in the RIOTS4 sample \citep{VargasSalazar20}, and is consistent with earlier suggestions that almost all field OB stars are ejected systems \citep{Gvaramadze12}.

Finally, our data strongly support the growing evidence for the post-mass-transfer model for the origin of classical Oe/Be stars \citep[e.g.,][]{Shao14, Boubert18, Klement19}. The kinematics and statistics for these objects are fully consistent with their origin as BSS ejection products, and are also consistent with those of the HMXBs in our sample. This model provides a simple and elegant explanation for the bimodality in the \vsini\ distribution and high, near-critical, Oe/Be rotation velocities. The close correspondence to BSS predicted frequencies also implies that {\bf Oe/Be disks are long-lived.}

\acknowledgments
We thank Jon Bjorkman, Karen Bjorkman, Matt Dallas, Kaitlin Kratter, Mathieu Renzo, Irene Vargas-Salazar, and the anonymous reviewer, for useful comments and discussions. We are also grateful to Sergio Sim\'on-D\'iaz for providing, and assisting us with, the {\sc iacob-broad} code. This work was funded by the National Science Foundation, grant AST-1514838 to MSO, and by the University of Michigan. M.M. acknowledges financial support from NASA grant ATP-170070. This research made use of Astropy, a community-developed core Python package for Astronomy \citep{Astropy}, as well as Numpy and Scipy \citep{Scipy}.

\break
\bibliographystyle{aasjournal}{}
\bibliography{ms}

\begin{thebibliography}{}
\expandafter\ifx\csname natexlab\endcsname\relax\def\natexlab#1{#1}\fi
\providecommand{\url}[1]{\href{#1}{#1}}

\bibitem[{{Astropy Collaboration} {et~al.}(2013){Astropy Collaboration},
  {Robitaille}, {Tollerud}, {Greenfield}, {Droettboom}, {Bray}, {Aldcroft},
  {Davis}, {Ginsburg}, {Price-Whelan}, {Kerzendorf}, {Conley}, {Crighton},
  {Barbary}, {Muna}, {Ferguson}, {Grollier}, {Parikh}, {Nair}, {Unther},
  {Deil}, {Woillez}, {Conseil}, {Kramer}, {Turner}, {Singer}, {Fox}, {Weaver},
  {Zabalza}, {Edwards}, {Azalee Bostroem}, {Burke}, {Casey}, {Crawford},
  {Dencheva}, {Ely}, {Jenness}, {Labrie}, {Lim}, {Pierfederici}, {Pontzen},
  {Ptak}, {Refsdal}, {Servillat}, \& {Streicher}}]{Astropy}
{Astropy Collaboration}, {Robitaille}, T.~P., {Tollerud}, E.~J., {et~al.} 2013,
  \aap, 558, A33

\bibitem[{{Banerjee} {et~al.}(2012){Banerjee}, {Kroupa}, \& {Oh}}]{Banerjee}
{Banerjee}, S., {Kroupa}, P., \& {Oh}, S. 2012, \apj, 746, 15

\bibitem[{{Blaauw}(1961)}]{Blaauw61}
{Blaauw}, A. 1961, \bain, 15, 265

\bibitem[{{Blaauw}(1993)}]{Blaauw93}
{Blaauw}, A. 1993, in Astronomical Society of the Pacific Conference Series,
  Vol.~35, Massive Stars: Their Lives in the Interstellar Medium, ed. J.~P.
  {Cassinelli} \& E.~B. {Churchwell}, 207

\bibitem[{{Bodensteiner} {et~al.}(2020){Bodensteiner}, {Shenar}, \&
  {Sana}}]{Bodensteiner20}
{Bodensteiner}, J., {Shenar}, T., \& {Sana}, H. 2020, \aap, 641, A42

\bibitem[{{Boubert} \& {Evans}(2018)}]{Boubert18}
{Boubert}, D., \& {Evans}, N.~W. 2018, \mnras, 477, 5261

\bibitem[{{Bragan{\c{c}}a} {et~al.}(2012){Bragan{\c{c}}a}, {Daflon}, {Cunha},
  {Bensby}, {Oey}, \& {Walth}}]{Braganca12}
{Bragan{\c{c}}a}, G.~A., {Daflon}, S., {Cunha}, K., {et~al.} 2012, \aj, 144,
  130

\bibitem[{{Brandt} \& {Podsiadlowski}(1995)}]{Brandt95}
{Brandt}, N., \& {Podsiadlowski}, P. 1995, \mnras, 274, 461

\bibitem[{{Brott} {et~al.}(2011){Brott}, {de Mink}, {Cantiello}, {Langer}, {de
  Koter}, {Evans}, {Hunter}, {Trundle}, \& {Vink}}]{Brott}
{Brott}, I., {de Mink}, S.~E., {Cantiello}, M., {et~al.} 2011, \aap, 530, A115

\bibitem[{{Chini} {et~al.}(2012){Chini}, {Hoffmeister}, {Nasseri}, {Stahl}, \&
  {Zinnecker}}]{Chini12}
{Chini}, R., {Hoffmeister}, V.~H., {Nasseri}, A., {Stahl}, O., \& {Zinnecker},
  H. 2012, \mnras, 424, 1925

\bibitem[{{Clarke} \& {Pringle}(1992)}]{Clarke92}
{Clarke}, C.~J., \& {Pringle}, J.~E. 1992, \mnras, 255, 423

\bibitem[{{Coe}(2005)}]{Coe05}
{Coe}, M.~J. 2005, \mnras, 358, 1379

\bibitem[{{Crowther}(1997)}]{Crowther}
{Crowther}, P.~A. 1997, in IAU Symposium, Vol. 189, IAU Symposium, ed. T.~R.
  {Bedding}, A.~J. {Booth}, \& J.~{Davis}, 137--146

\bibitem[{{de Mink} {et~al.}(2009){de Mink}, {Cantiello}, {Langer}, {Pols},
  {Brott}, \& {Yoon}}]{deMink09}
{de Mink}, S.~E., {Cantiello}, M., {Langer}, N., {et~al.} 2009, \aap, 497, 243

\bibitem[{{de Mink} {et~al.}(2013){de Mink}, {Langer}, {Izzard}, {Sana}, \& {de
  Koter}}]{deMink13}
{de Mink}, S.~E., {Langer}, N., {Izzard}, R.~G., {Sana}, H., \& {de Koter}, A.
  2013, \apj, 764, 166

\bibitem[{{de Wit} {et~al.}(2005){de Wit}, {Testi}, {Palla}, \&
  {Zinnecker}}]{deWit05}
{de Wit}, W.~J., {Testi}, L., {Palla}, F., \& {Zinnecker}, H. 2005, \aap, 437,
  247

\bibitem[{{Dufton} {et~al.}(2013){Dufton}, {Langer}, {Dunstall}, {Evans},
  {Brott}, {de Mink}, {Howarth}, {Kennedy}, {McEvoy}, {Potter},
  {Ram{\'\i}rez-Agudelo}, {Sana}, {Sim{\'o}n-D{\'\i}az}, {Taylor}, \&
  {Vink}}]{Dufton13}
{Dufton}, P.~L., {Langer}, N., {Dunstall}, P.~R., {et~al.} 2013, \aap, 550,
  A109

\bibitem[{{Ekstr{\"o}m} {et~al.}(2008){Ekstr{\"o}m}, {Meynet}, {Maeder}, \&
  {Barblan}}]{Ekstrom08}
{Ekstr{\"o}m}, S., {Meynet}, G., {Maeder}, A., \& {Barblan}, F. 2008, \aap,
  478, 467

\bibitem[{{Evans} {et~al.}(2004){Evans}, {Howarth}, {Irwin}, {Burnley}, \&
  {Harries}}]{Evans04}
{Evans}, C.~J., {Howarth}, I.~D., {Irwin}, M.~J., {Burnley}, A.~W., \&
  {Harries}, T.~J. 2004, \mnras, 353, 601

\bibitem[{{Fujii} \& {Portegies Zwart}(2011)}]{FujiiPZ}
{Fujii}, M.~S., \& {Portegies Zwart}, S. 2011, Science, 334, 1380

\bibitem[{{Fujii} \& {Portegies Zwart}(2014)}]{Fujii14}
---. 2014, \mnras, 439, 1003

\bibitem[{{Gies}(1987)}]{Gies87}
{Gies}, D.~R. 1987, \apjs, 64, 545

\bibitem[{{Gies} \& {Bolton}(1986)}]{Gies86}
{Gies}, D.~R., \& {Bolton}, C.~T. 1986, \apjs, 61, 419

\bibitem[{{Golden-Marx} {et~al.}(2016){Golden-Marx}, {Oey}, {Lamb}, {Graus}, \&
  {White}}]{Golden-Marx16}
{Golden-Marx}, J.~B., {Oey}, M.~S., {Lamb}, J.~B., {Graus}, A.~S., \& {White},
  A.~S. 2016, \apj, 819, 55

\bibitem[{{Gott}(1971)}]{Gott71}
{Gott}, J.~R. 1971, \nat, 234, 342

\bibitem[{{Graus} {et~al.}(2012){Graus}, {Lamb}, \& {Oey}}]{Graus12}
{Graus}, A.~S., {Lamb}, J.~B., \& {Oey}, M.~S. 2012, \apj, 759, 10

\bibitem[{{Gvaramadze} {et~al.}(2012){Gvaramadze}, {Weidner}, {Kroupa}, \&
  {Pflamm-Altenburg}}]{Gvaramadze12}
{Gvaramadze}, V.~V., {Weidner}, C., {Kroupa}, P., \& {Pflamm-Altenburg}, J.
  2012, \mnras, 424, 3037

\bibitem[{{Haberl} \& {Sturm}(2016)}]{HaberlSturm}
{Haberl}, F., \& {Sturm}, R. 2016, \aap, 586, A81

\bibitem[{{Harries} {et~al.}(2003){Harries}, {Hilditch}, \&
  {Howarth}}]{Harries03}
{Harries}, T.~J., {Hilditch}, R.~W., \& {Howarth}, I.~D. 2003, \mnras, 339, 157

\bibitem[{{Hoogerwerf} {et~al.}(2000){Hoogerwerf}, {de Bruijne}, \& {de
  Zeeuw}}]{Hoogerwerf00}
{Hoogerwerf}, R., {de Bruijne}, J.~H.~J., \& {de Zeeuw}, P.~T. 2000, \apjl,
  544, L133

\bibitem[{{Hoogerwerf} {et~al.}(2001){Hoogerwerf}, {de Bruijne}, \& {de
  Zeeuw}}]{Hoogerwerf01}
---. 2001, \aap, 365, 49

\bibitem[{{Klement} {et~al.}(2019){Klement}, {Carciofi}, {Rivinius}, {Ignace},
  {Matthews}, {Torstensson}, {Gies}, {Vieira}, {Richardson}, {Domiciano de
  Souza}, {Bjorkman}, {Hallinan}, {Faes}, {Mota}, {Gullingsrud}, {de Breuck},
  {Kervella}, {Cur{\'e}}, \& {Gunawan}}]{Klement19}
{Klement}, R., {Carciofi}, A.~C., {Rivinius}, T., {et~al.} 2019, \apj, 885, 147

\bibitem[{{Kochanek} {et~al.}(2019){Kochanek}, {Auchettl}, \&
  {Belczynski}}]{Kochanek19}
{Kochanek}, C.~S., {Auchettl}, K., \& {Belczynski}, K. 2019, \mnras, 485, 5394

\bibitem[{{Lada} \& {Lada}(2003)}]{LadaLada03}
{Lada}, C.~J., \& {Lada}, E.~A. 2003, \araa, 41, 57

\bibitem[{{Lamb} {et~al.}(2013){Lamb}, {Oey}, {Graus}, {Adams}, \&
  {Segura-Cox}}]{Lamb13}
{Lamb}, J.~B., {Oey}, M.~S., {Graus}, A.~S., {Adams}, F.~C., \& {Segura-Cox},
  D.~M. 2013, \apj, 763, 101

\bibitem[{{Lamb} {et~al.}(2016){Lamb}, {Oey}, {Segura-Cox}, {Graus}, {Kiminki},
  {Golden-Marx}, \& {Parker}}]{Lamb16}
{Lamb}, J.~B., {Oey}, M.~S., {Segura-Cox}, D.~M., {et~al.} 2016, \apj, 817, 113

\bibitem[{{Lamb} {et~al.}(2010){Lamb}, {Oey}, {Werk}, \& {Ingleby}}]{Lamb10}
{Lamb}, J.~B., {Oey}, M.~S., {Werk}, J.~K., \& {Ingleby}, L.~D. 2010, \apj,
  725, 1886

\bibitem[{{Leonard} \& {Duncan}(1988)}]{Leonard88}
{Leonard}, P. J.~T., \& {Duncan}, M.~J. 1988, \aj, 96, 222

\bibitem[{{Maeder} \& {Meynet}(2000)}]{Maeder00}
{Maeder}, A., \& {Meynet}, G. 2000, \araa, 38, 143

\bibitem[{{Ma{\'\i}z Apell{\'a}niz} {et~al.}(2018){Ma{\'\i}z Apell{\'a}niz},
  {Pantaleoni Gonz{\'a}lez}, {Barb{\'a}}, {Sim{\'o}n-D{\'\i}az}, {Negueruela},
  {Lennon}, {Sota}, \& {Trigueros P{\'a}ez}}]{MaizApellaniz}
{Ma{\'\i}z Apell{\'a}niz}, J., {Pantaleoni Gonz{\'a}lez}, M., {Barb{\'a}},
  R.~H., {et~al.} 2018, \aap, 616, A149

\bibitem[{{Mason} {et~al.}(2009){Mason}, {Hartkopf}, {Gies}, {Henry}, \&
  {Helsel}}]{Mason09}
{Mason}, B.~D., {Hartkopf}, W.~I., {Gies}, D.~R., {Henry}, T.~J., \& {Helsel},
  J.~W. 2009, \aj, 137, 3358

\bibitem[{{Massey}(2002)}]{Massey02}
{Massey}, P. 2002, \apjs, 141, 81

\bibitem[{{Massey} {et~al.}(2005){Massey}, {Puls}, {Pauldrach}, {Bresolin},
  {Kudritzki}, \& {Simon}}]{Massey05}
{Massey}, P., {Puls}, J., {Pauldrach}, A.~W.~A., {et~al.} 2005, \apj, 627, 477

\bibitem[{{McKee} \& {Ostriker}(2007)}]{McKee07}
{McKee}, C.~F., \& {Ostriker}, E.~C. 2007, \araa, 45, 565

\bibitem[{{McSwain} \& {Gies}(2005)}]{McSwain05}
{McSwain}, M.~V., \& {Gies}, D.~R. 2005, \apjs, 161, 118

\bibitem[{{Moe} \& {Di Stefano}(2017)}]{Moe17}
{Moe}, M., \& {Di Stefano}, R. 2017, \apjs, 230, 15

\bibitem[{{Oey} {et~al.}(2004){Oey}, {King}, \& {Parker}}]{Oey04}
{Oey}, M.~S., {King}, N.~L., \& {Parker}, J.~W. 2004, \aj, 127, 1632

\bibitem[{{Oey} {et~al.}(2013){Oey}, {Lamb}, {Kushner}, {Pellegrini}, \&
  {Graus}}]{Oey2013}
{Oey}, M.~S., {Lamb}, J.~B., {Kushner}, C.~T., {Pellegrini}, E.~W., \& {Graus},
  A.~S. 2013, \apj, 768, 66

\bibitem[{{Oey} {et~al.}(2018){Oey}, {Dorigo Jones}, {Castro}, {Zivick},
  {Besla}, {Januszewski}, {Moe}, {Kallivayalil}, \& {Lennon}}]{PaperI}
{Oey}, M.~S., {Dorigo Jones}, J., {Castro}, N., {et~al.} 2018, \apjl, 867, L8

\bibitem[{{Oh} \& {Kroupa}(2016)}]{OhKroupa}
{Oh}, S., \& {Kroupa}, P. 2016, \aap, 590, A107

\bibitem[{{Oh} {et~al.}(2015){Oh}, {Kroupa}, \&
  {Pflamm-Altenburg}}]{OhKroupa15}
{Oh}, S., {Kroupa}, P., \& {Pflamm-Altenburg}, J. 2015, \apj, 805, 92

\bibitem[{{Parker} \& {Goodwin}(2007)}]{Parker07}
{Parker}, R.~J., \& {Goodwin}, S.~P. 2007, \mnras, 380, 1271

\bibitem[{{Pawlak} {et~al.}(2013){Pawlak}, {Graczyk}, {Soszy{\'n}ski},
  {Pietrukowicz}, {Poleski}, {Udalski}, {Szyma{\'n}ski}, {Kubiak},
  {Pietrzy{\'n}ski}, {Wyrzykowski}, {Ulaczyk}, {Koz{\l}owski}, \&
  {Skowron}}]{Pawlak13}
{Pawlak}, M., {Graczyk}, D., {Soszy{\'n}ski}, I., {et~al.} 2013, \actaa, 63,
  323

\bibitem[{{Pawlak} {et~al.}(2016){Pawlak}, {Soszy{\'n}ski}, {Udalski},
  {Szyma{\'n}ski}, {Wyrzykowski}, {Ulaczyk}, {Poleski}, {Pietrukowicz},
  {Koz{\l}owski}, {Skowron}, {Skowron}, {Mr{\'o}z}, \& {Hamanowicz}}]{Pawlak16}
{Pawlak}, M., {Soszy{\'n}ski}, I., {Udalski}, A., {et~al.} 2016, \actaa, 66,
  421

\bibitem[{{Perets} \& {{\v{S}}ubr}(2012)}]{PeretsSubr}
{Perets}, H.~B., \& {{\v{S}}ubr}, L. 2012, \apj, 751, 133

\bibitem[{{Pflamm-Altenburg} \& {Kroupa}(2010)}]{Pflamm-Altenburg10}
{Pflamm-Altenburg}, J., \& {Kroupa}, P. 2010, \mnras, 404, 1564

\bibitem[{{Pols} {et~al.}(1991){Pols}, {Cote}, {Waters}, \& {Heise}}]{Pols91}
{Pols}, O.~R., {Cote}, J., {Waters}, L.~B.~F.~M., \& {Heise}, J. 1991, \aap,
  241, 419

\bibitem[{{Poveda} {et~al.}(1967){Poveda}, {Ruiz}, \& {Allen}}]{Poveda67}
{Poveda}, A., {Ruiz}, J., \& {Allen}, C. 1967, Boletin de los Observatorios
  Tonantzintla y Tacubaya, 4, 86

\bibitem[{{Ramachandran} {et~al.}(2019){Ramachandran}, {Hamann}, {Oskinova},
  {Gallagher}, {Hainich}, {Shenar}, {Sand er}, {Todt}, \&
  {Fulmer}}]{Ramachandran2019}
{Ramachandran}, V., {Hamann}, W.~R., {Oskinova}, L.~M., {et~al.} 2019, \aap,
  625, A104

\bibitem[{{Rawls} {et~al.}(2011){Rawls}, {Orosz}, {McClintock}, {Torres},
  {Bailyn}, \& {Buxton}}]{Rawls2011}
{Rawls}, M.~L., {Orosz}, J.~A., {McClintock}, J.~E., {et~al.} 2011, \apj, 730,
  25

\bibitem[{{Renzo} {et~al.}(2019){Renzo}, {Zapartas}, {de Mink}, {G{\"o}tberg},
  {Justham}, {Farmer}, {Izzard}, {Toonen}, \& {Sana}}]{Renzo19}
{Renzo}, M., {Zapartas}, E., {de Mink}, S.~E., {et~al.} 2019, \aap, 624, A66

\bibitem[{{Rivinius} {et~al.}(2013){Rivinius}, {Baade}, {Townsend}, {Carciofi},
  \& {{\v{S}}tefl}}]{Rivinius13}
{Rivinius}, T., {Baade}, D., {Townsend}, R.~H.~D., {Carciofi}, A.~C., \&
  {{\v{S}}tefl}, S. 2013, \aap, 559, L4

\bibitem[{{Sana} {et~al.}(2012){Sana}, {de Mink}, {de Koter}, {Langer},
  {Evans}, {Gieles}, {Gosset}, {Izzard}, {Le Bouquin}, \& {Schneider}}]{Sana12}
{Sana}, H., {de Mink}, S.~E., {de Koter}, A., {et~al.} 2012, Science, 337, 444

\bibitem[{{Schoettler} {et~al.}(2020){Schoettler}, {de Bruijne}, {Vaher}, \&
  {Parker}}]{Schoettler20}
{Schoettler}, C., {de Bruijne}, J., {Vaher}, E., \& {Parker}, R.~J. 2020,
  \mnras, 495, 3104

\bibitem[{{Shao} \& {Li}(2014)}]{Shao14}
{Shao}, Y., \& {Li}, X.-D. 2014, \apj, 796, 37

\bibitem[{{Sim{\'o}n-D{\'\i}az} \& {Herrero}(2014)}]{Diaz14}
{Sim{\'o}n-D{\'\i}az}, S., \& {Herrero}, A. 2014, \aap, 562, A135

\bibitem[{{Stone}(1991)}]{Stone91}
{Stone}, R.~C. 1991, \aj, 102, 333

\bibitem[{{van Bever} \& {Vanbeveren}(1997)}]{VanBever97}
{van Bever}, J., \& {Vanbeveren}, D. 1997, \aap, 322, 116

\bibitem[{{van den Heuvel}(1981)}]{Heuvel81}
{van den Heuvel}, E.~P.~J. 1981, in IAU Symposium, Vol.~95, Pulsars: 13 Years
  of Research on Neutron Stars, ed. W.~{Sieber} \& R.~{Wielebinski}, 379--394

\bibitem[{{Vargas-Salazar} {et~al.}(2020){Vargas-Salazar}, {Oey}, {Barnes},
  {Chen}, {Castro}, {Kratter}, \& {Faerber}}]{VargasSalazar20}
{Vargas-Salazar}, I., {Oey}, M.~S., {Barnes}, J.~R., {et~al.} 2020, \apj,
  submitted

\bibitem[{{Vinciguerra} {et~al.}(2020){Vinciguerra}, {Neijssel},
  {Vigna-G{\'o}mez}, {Mandel}, {Podsiadlowski}, {Maccarone}, {Nicholl},
  {Kingdon}, {Perry}, \& {Salemi}}]{Vinciguerra2020}
{Vinciguerra}, S., {Neijssel}, C.~J., {Vigna-G{\'o}mez}, A., {et~al.} 2020,
  arXiv e-prints, arXiv:2003.00195

\bibitem[{{Virtanen} {et~al.}(2020){Virtanen}, {Gommers}, {Oliphant},
  {Haberland}, {Reddy}, {Cournapeau}, {Burovski}, {Peterson}, {Weckesser},
  {Bright}, {van der Walt}, {Brett}, {Wilson}, {Jarrod Millman}, {Mayorov},
  {Nelson}, {Jones}, {Kern}, {Larson}, {Carey}, {Polat}, {Feng}, {Moore}, {Vand
  erPlas}, {Laxalde}, {Perktold}, {Cimrman}, {Henriksen}, {Quintero}, {Harris},
  {Archibald}, {Ribeiro}, {Pedregosa}, {van Mulbregt}, \&
  {Contributors}}]{Scipy}
{Virtanen}, P., {Gommers}, R., {Oliphant}, T.~E., {et~al.} 2020, Nature
  Methods, 17, 261

\bibitem[{{Walborn} {et~al.}(2014){Walborn}, {Sana}, {Sim{\'o}n-D{\'\i}az},
  {Ma{\'\i}z Apell{\'a}niz}, {Taylor}, {Evans}, {Markova}, {Lennon}, \& {de
  Koter}}]{Walborn14}
{Walborn}, N.~R., {Sana}, H., {Sim{\'o}n-D{\'\i}az}, S., {et~al.} 2014, \aap,
  564, A40

\bibitem[{{Wang} {et~al.}(2020){Wang}, {Langer}, {Schootemeijer}, {Castro},
  {Adscheid}, {Marchant}, \& {Hastings}}]{Wang2020}
{Wang}, C., {Langer}, N., {Schootemeijer}, A., {et~al.} 2020, \apjl, 888, L12

\bibitem[{{Wolff} {et~al.}(2007){Wolff}, {Strom}, {Dror}, \& {Venn}}]{Wolff07}
{Wolff}, S.~C., {Strom}, S.~E., {Dror}, D., \& {Venn}, K. 2007, \aj, 133, 1092

\bibitem[{{Zaritsky} {et~al.}(2002){Zaritsky}, {Harris}, {Thompson}, {Grebel},
  \& {Massey}}]{Zaritsky02}
{Zaritsky}, D., {Harris}, J., {Thompson}, I.~B., {Grebel}, E.~K., \& {Massey},
  P. 2002, \aj, 123, 855

\bibitem[{{Zwicky}(1957)}]{Zwicky}
{Zwicky}, F. 1957, {Morphological astronomy}

\end{thebibliography}
\end{document}